\documentclass[reprint,superscriptaddress,amsmath,amssymb,aps,pra]{revtex4-1}

\usepackage{graphicx}
\usepackage{hyperref}
\usepackage{subcaption}
\usepackage[english]{babel}

\begin{document}

\title{Efficient, high-resolution resonance laser ionization spectroscopy using weak transitions to long-lived excited states}

\author{R.~P.~de Groote}
\email{ruben.degroote@fys.kuleuven.be}
\affiliation{KU Leuven, Instituut voor Kern- en Stralingsfysica, B-3001 Leuven, Belgium}
\author{M.~Verlinde}
\affiliation{KU Leuven, Instituut voor Kern- en Stralingsfysica, B-3001 Leuven, Belgium}
\author{V.~Sonnenschein}
\affiliation{Department of Physics, University of Jyv\"{a}skyl\"{a}, P.O. Box 35, 40014 Jyv\"{a}skyl\"{a}, Finland}
\author{K.~T.~Flanagan}
\affiliation{School of Physics and Astronomy, The University of Manchester, Manchester M13 9PL, UK}
\author{I.~Moore}
\affiliation{Department of Physics, University of Jyv\"{a}skyl\"{a}, P.O. Box 35, 40014 Jyv\"{a}skyl\"{a}, Finland}
\author{G.~Neyens}
\affiliation{KU Leuven, Instituut voor Kern- en Stralingsfysica, B-3001 Leuven, Belgium}

\begin{abstract}
Laser spectroscopic studies on minute samples of exotic radioactive nuclei require very efficient experimental techniques. In addition, high resolving powers are required to allow extraction of nuclear structure information. Here we demonstrate that by using weak atomic transitions, resonance laser ionization spectroscopy is achieved with the required high efficiency (1-10\%) and precision (linewidths of tens of MHz). We illustrate experimentally and through the use of simulations how the narrow experimental linewidths are achieved and how distorted resonance ionization spectroscopy lineshapes can be avoided. The role of the delay of the ionization laser pulse with respect to the excitation laser pulse is crucial: the use of a delayed ionization step permits the best resolving powers and lineshapes.  A high efficiency is maintained if the intermediate level has a lifetime that is at least of the order of the excitation laser pulse width. A model that describes this process reproduces well the observed features and will help to optimize the conditions for future experiments \footnote{Simulation code available upon request to the authors.}. 
\end{abstract}

\maketitle

\section{Introduction} \label{sec:intro}

The study of the fundamental ground-state properties of exotic nuclei is one of the challenges in contemporary nuclear physics research \cite{nupecc2010}. Laser spectrosopic methods contribute to this ongoing research endeavor by providing information on the nuclear electromagnetic moments, spins, and changes in mean-squared charge radii. These observables provide key input towards a theoretical description of the nucleus, as illustrated e.g. in \cite{ruiz2016}. Experimental measurements on very exotic nuclei are challenging, reflecting a combination of short half-lives and production in only minute quantities, accompanied by a large amount of unwanted contamination. Furthermore, facilities that produce these exotic nuclei only allot a limited time to a given experiment, which means that the study of the most exotic cases requires techniques that are both very selective and efficient. In addition, such measurements are often performed only once, so systematic uncertainties must be understood, removed, or at least minimized. 

Many laser spectroscopic techniques have been applied in nuclear physics research, each with their strengths and weaknesses \cite{Cheal2010,Blaum2013,Campbell2016}. Resonance Ionization Spectroscopy (RIS) methods, which rely on multi-step laser ionization and subsequent ion detection, are typically very sensitive, motivating the development of numerous RIS experiments at online isotope separators \cite{Fedosseev2012}. To achieve efficient laser ionization, pulsed laser systems are typically used, often operating at high powers. This leads to a drawback for spectroscopy, including unwanted lineshape distortion and/or line broadening. These effects can become apparent when performing high-resolution RIS, as illustrated recently in \cite{de_Groote_2015}. Given the current developments towards high-resolution RIS of exotic nuclei in e.g. collinear RIS \cite{de_Groote_2015} and in-gas-jet laser spectroscopy \cite{Kudryavtsev2013,Raeder2016}, a detailed understanding of the interaction of atoms with pulsed lasers is vital. 

This article will present a model that describes the RIS process using CW or pulsed lasers for the resonant excitation and non-resonant ionization step (section \ref{sec:model}). Spontaneous decay of the intermediate level is taken into account and time delays between the two excitation steps are investigated. Through both model simulations and experimental verification, this article will address how some of the aforementioned detrimental line distortion effects can be understood and avoided by delaying the ionization laser pulse. This will be presented in section \ref{sec:dist}. Furthermore, through the same delayed-ionization approach, it is possible to remove virtually all power broadening due to both lasers in a two-step RIS scheme, which could be important for future high-precision studies on radioactive isotopes. This will be illustrated experimentally and through simulations in section \ref{sec:pb}. 

The delayed ionization method is greatly enhanced by using a weak transition to a long-lived excited state. Firstly, with short-lived excited states, a significant fraction of the excited state population would decay before the ionization can take place, reducing the efficiency of the method. Secondly, long-lived states result in intrinsically narrower linewidth since their natural width is smaller. The feasibility of using weak transitions for efficient resonance laser ionization spectroscopy using both continuous wave and pulsed lasers will be addressed in section \ref{sec:eff}.

\section{A Model for RIS}\label{sec:model}

The evolution of the population of a two level system irradiated by a laser tuned close to resonance can be calculated by solving the Schr\"odinger equations with the following Hamiltonian:
\begin{align}
H &= \frac{\hbar}{2} \begin{pmatrix} 0 & \Omega(t)\\ \Omega(t)& 2\Delta \end{pmatrix},
\end{align}
where $\Omega(t)$ is the coupling parameter of the two states, also called the Rabi frequency, and $\Delta$ is the laser-atom detuning. Defining the laser frequency as $\omega_e$, the ground state level energy $\hbar\omega_0=0$ and the excited state energy $\hbar\omega_1$,  $\Delta = \omega_1 - \omega_e$. When using linearly polarized light, the coupling parameter $\Omega(t)$ can be calculated using
\begin{align}
    \Omega =& \sqrt{A P_e} \left(\frac{c}{\omega_1 - \omega_0}\right)^{3/2} (2F_1+1) \notag\\
    & \times \sum_{m_{F_0},m_{F_1}}\begin{pmatrix} F_1 & 1 & F_0 \\ -m_{F_1} &0&m_{F_0}\end{pmatrix} \begin{Bmatrix} J_1 & F_1 & I \\ F_0 & J_0 & 1\end{Bmatrix},\label{rabi_frequency}
\end{align}
with $F_i$ the total angular momentum of state $i$, $()$ and $\{\}$ respectively Wigner 3J and 6J symbols, $A$ the Einstein $A$ coefficient of the transition and $P_e(t)$ the power of the laser. 

A second laser with laser power $P_i(t)$ can ionize excited atoms at a rate of $\Gamma = P_i(t)\sigma$, with $\sigma$ the non-resonant photo-ionization cross section of the excited state at the wavelength of the ionization laser. Photo-ionization requires modeling population loss, since population has to flow out of the two-level system, into the continuum. This requires a non-Hermitian Hamiltonian, given by
\begin{align}
H &= \frac{\hbar}{2} \begin{pmatrix} 0 & \Omega(t)\\ \Omega(t)& 2\Delta + 2 S(t) - i\Gamma(t) \end{pmatrix},\label{eq:ionization_ham}
\end{align}
In this Hamiltonian $S$ is the net dynamic Stark shift induced by both the ionization laser and excitation laser, with each laser contributing a shift proportional to the laser power \cite{Delone_1999,kumekov1981dynamic}. If there are no relaxation processes, the evolution of the system can be calculated using the time-dependent Schr\"odinger equation
\begin{align}
    \dot{\rho} = \frac{1}{i\hbar}\left(H\rho - \rho H^\dagger\right).\label{eq:density_equation}
\end{align}
The right-hand side of this equation reduces to the more familiar commutator $[H, \rho]$ for Hermitian $H$. The populations of the hyperfine levels are the diagonal elements of the density matrix, $\rho_{ii}(t)$. Usually the spontaneous decay of the excited state to the ground state cannot be neglected. Including incoherent relaxation processes into equation \eqref{eq:density_equation} can be done as follows:
\begin{align}
    \dot{\rho} = \frac{1}{i\hbar}\left(H\rho - \rho H^\dagger\right) + L(\rho). \label{eq:full_eom}
\end{align}
For the simplified two-level system, $L$ is given by
\begin{align}
L &= \begin{pmatrix} A\rho_{11} &  -\frac12 A\rho_{01}\\ -\frac12 A\rho_{10} & -A\rho_{11} \end{pmatrix},
\end{align}
This additional term in the equations of motion causes decay of the excited state to the lower-lying state, and exponentially dampens the coherence terms. \\

\begin{figure}[ht!]
\begin{center}
\includegraphics[width=\columnwidth]{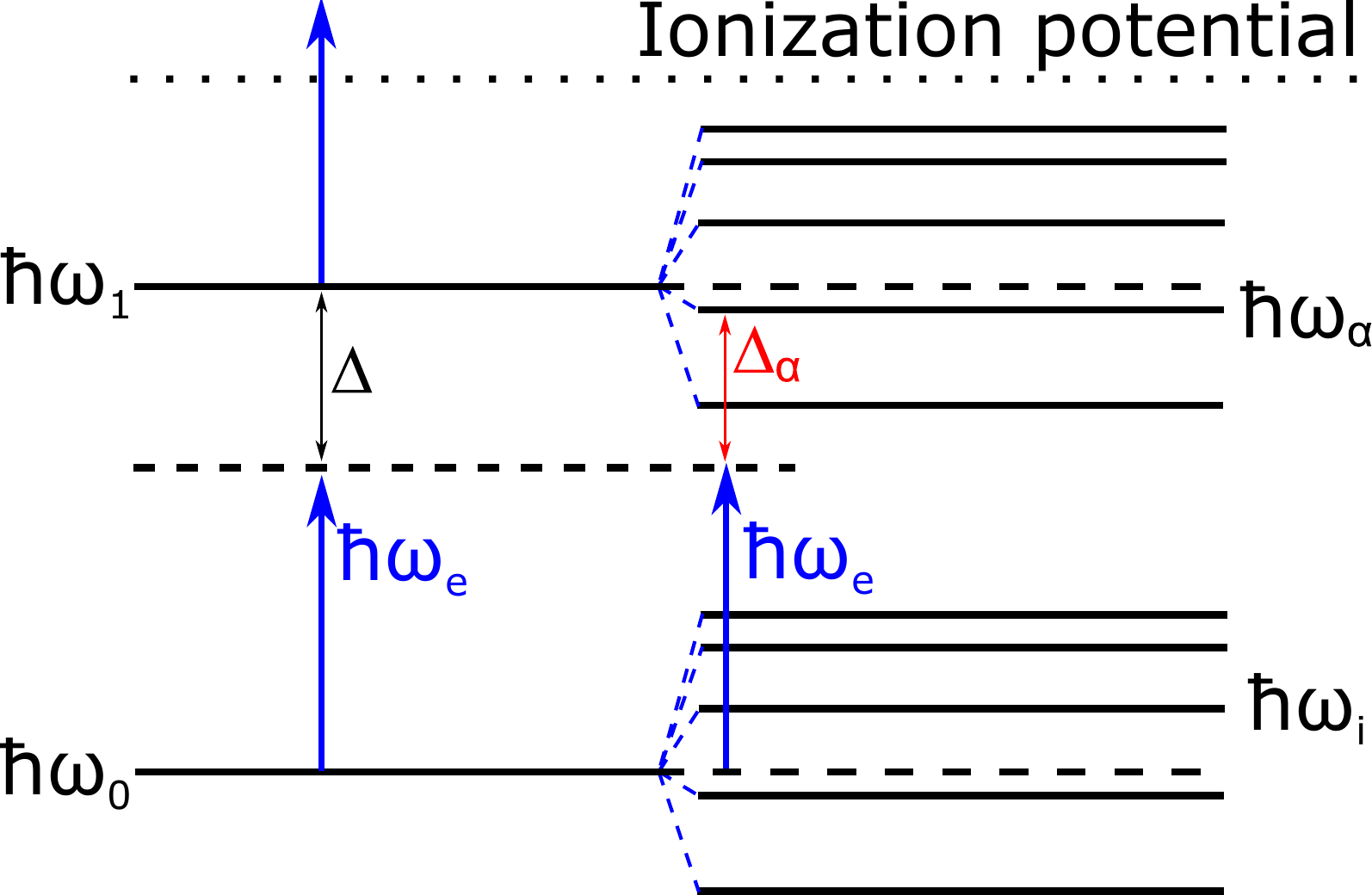}
\caption{An atom with several hyperfine levels in the ground-state and in the excited state multiplet.  Note the figure is not to scale, since the hyperfine splitting is typically $10^6$ times smaller than the transition energy.}
\label{fig:general_picture}
\end{center}
\end{figure}

These equations of motion can be generalized to systems with multiple hyperfine levels in a ground- or excited state multiplet (see Fig. \ref{fig:general_picture}) in a relatively straightforward manner. Using Greek indices for levels in the excited state multiplet and roman indices for the ground state multiplet, the Hamiltonian for such a system can be written down as
\begin{align}
&H_{ii}(t) = \hbar \omega_i \\
&H_{ij}(t) = 0 \\
&H_{\alpha\alpha}(t) = \hbar (S_{\alpha}(t) + \Delta_{\alpha} - \frac{i}{2} \Gamma(t) ) \\
&H_{i\alpha}(t) =  H_{\alpha i}(t) = \frac{\hbar}{2} \Omega_{i\alpha}(t)\\
&H_{\alpha\beta}(t) =  - \frac{\hbar}{2} \Gamma(t)(q+i), \label{eq:offdiag}
\end{align}
with $\hbar\omega_i$ and $\hbar\omega_\alpha$ the energy of the atomic states of the ground- and excited-state hyperfine multiplet, and $\Delta_\alpha = \omega_\alpha - \omega_e$. The off-diagonal terms in \eqref{eq:offdiag} are due to the embedding of structure into the continuum by the high-power ionization laser, and are characterized by a Fano parameter $q$ \cite{Knight1990}. This $q$ parameter plays a role in laser-induced continuum phenomena and can be calculated from first principles (see e.g. \cite{Dai1987,Nakajima1994,Yatsenko1997,Yatsenko1999}). It induces a coupling between the excited state multiplet levels by the ionization laser via the interaction with the continuum. This coupling can influence of RIS line profiles, but only for either very large values of $q$ or for unrealistically high laser powers. For the purpose of this article the Fano parameter will be taken to be zero.

The generalized form of the matrix $L$ can be written down using the partial decay rates defined as 
\begin{align}
\gamma_{\alpha i} & = \frac{4 \alpha}{3} \frac{| \omega_\alpha - \omega_i |^3}{c^2} (2F_i+1)(2J_\alpha+1)(2J_i+1)\\ 
& \times |\left\langle \alpha,L_\alpha || r || i,L_i\right\rangle|^2 \begin{Bmatrix} J_\alpha & 1 & J_i \\ F_i & I & F_\alpha  \end{Bmatrix}^2 \begin{Bmatrix} L_\alpha & 1 & L_i \\ J_i & S & J_\alpha  \end{Bmatrix}^2. \notag
\end{align}
These rates can be calculated using the observation that the partial decay rates of an excited hyperfine level should sum up to the Einstein $A$ coefficient. Using this definition of the partial rates, $L$ can be written down as 
\begin{align}
&L(\rho)_{ii} = \sum_\alpha \rho_{\alpha\alpha} \gamma_{\alpha i} \\
&L(\rho)_{\alpha\alpha} = - \sum_i \rho_{\alpha\alpha} \gamma_{\alpha i} \\
&L(\rho)_{\alpha i} = - \frac{\rho_{\alpha i}}{2} \sum_j \gamma_{\alpha j} \\
&L(\rho)_{i\alpha} = - \frac{\rho_{i \alpha}}{2} \sum_j \gamma_{\alpha j} \\
&L(\rho)_{\alpha \beta} = - \frac{\rho_{\alpha\beta}}{2} \sum_j \gamma_{\alpha j} + \gamma_{\beta j}.
\end{align}

These are the generalized equations that will be used for the simulations presented throughout this article. The computer code developed in Python used to run the simulations, is available upon request to the authors.

\section{Power broadening and delayed ionization}\label{sec:pb}

\subsection{Model Predictions}

\begin{figure}[ht!]
\begin{center}
\includegraphics[width=\columnwidth]{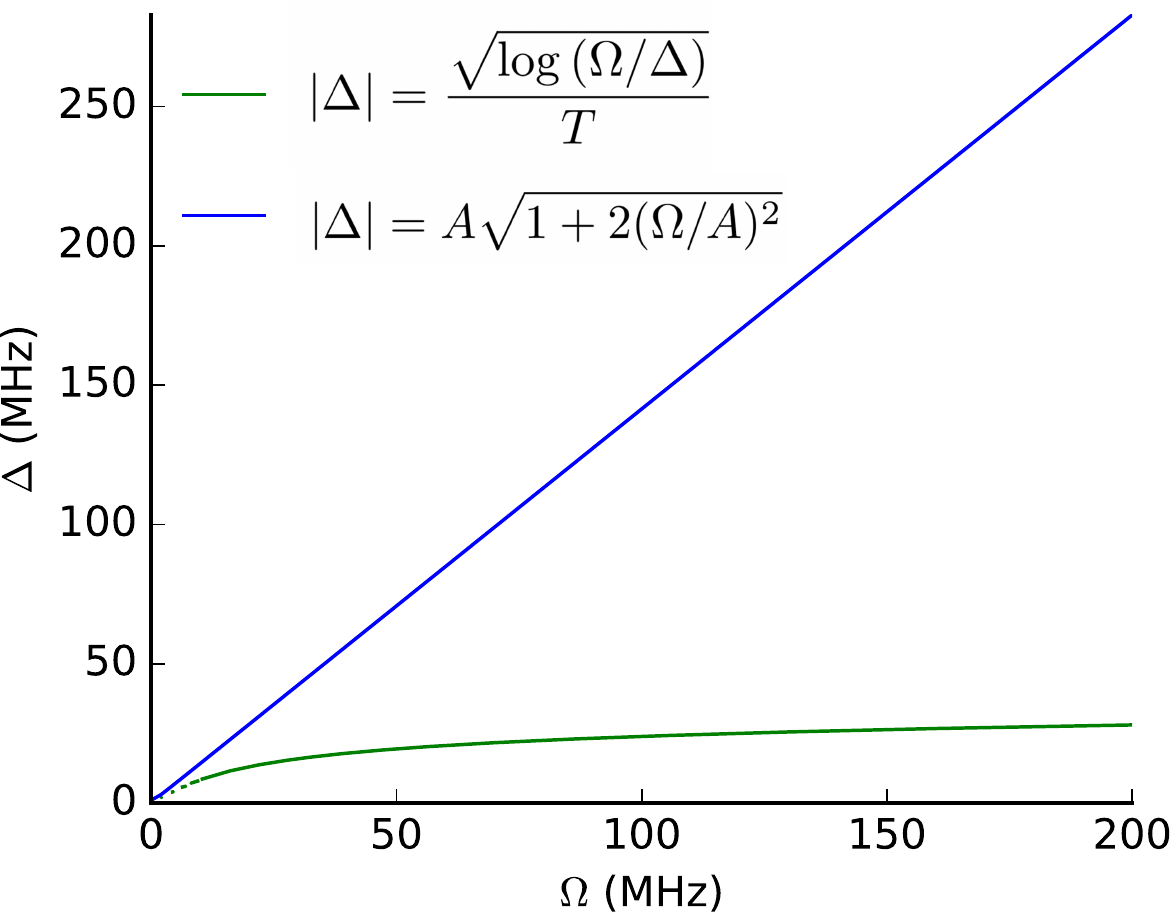}
\caption{Comparison of equations \eqref{eq:power_broadening} and \eqref{eq:Gaussbroadening},  representing the line broadening $\Delta$ due to the excitation coupling $\Omega$, for an Einstein A-coefficient of 1\,MHz for continuous-wave lasers and T=50\,ns for pulsed lasers. }
\label{fig:pb_eqs}
\end{center}
\end{figure}

\begin{figure*}[ht!]
    \centering
    \begin{subfigure}[b]{0.485\textwidth}
        \centering
        \includegraphics[width=\textwidth]{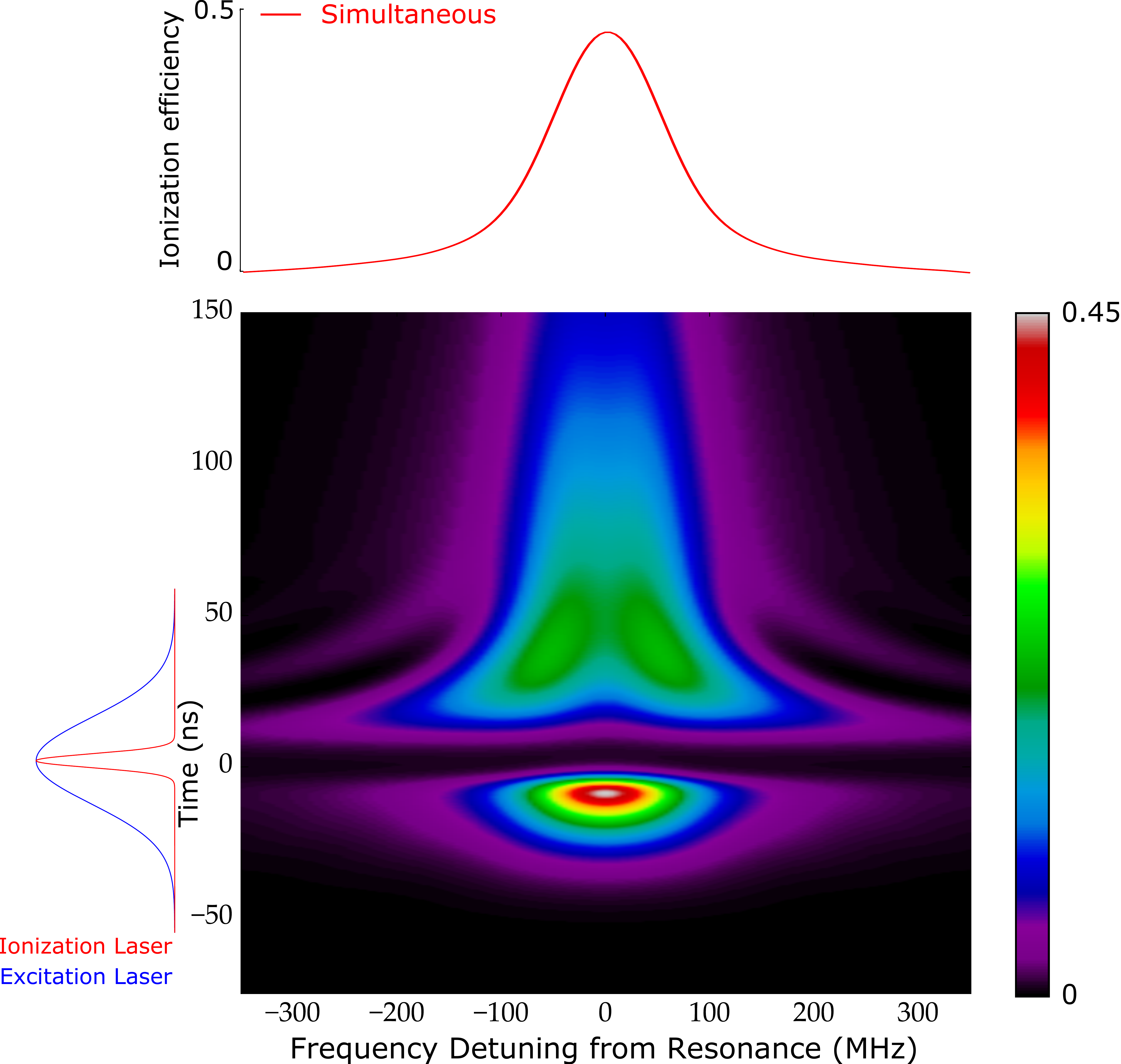}
        \caption[]%
        {{Excited state population when using simultaneous laser pulses (see diagram on the left). The resonance ionization spectrum (red curve in the top panel) is power broadened by both laser pulses.}}
        \label{fig:2d1}
    \end{subfigure}
    \hfill
    \begin{subfigure}[b]{0.485\textwidth}  
        \centering 
        \includegraphics[width=\textwidth]{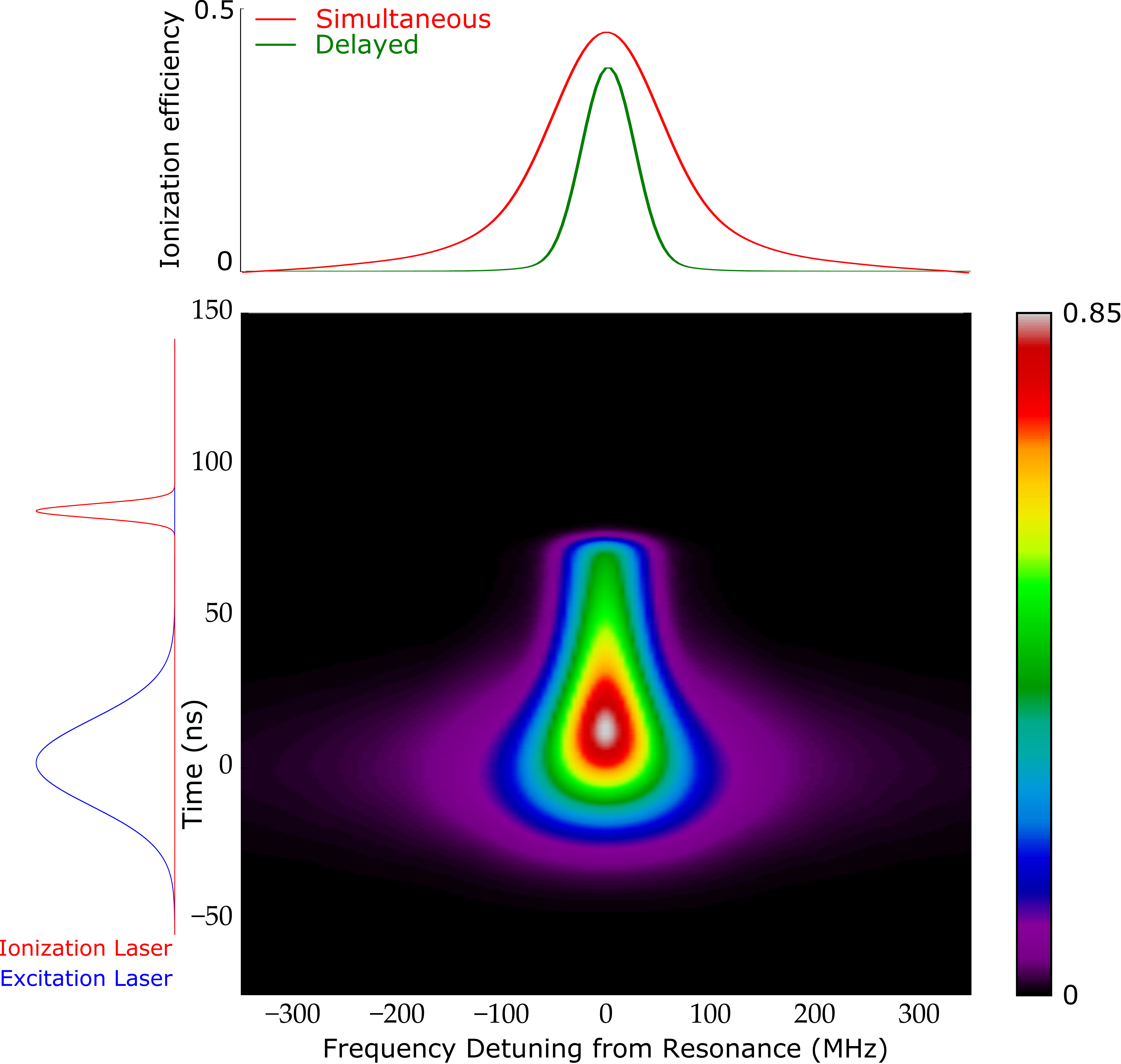}
        \caption[]%
        {{As Fig. \ref{fig:2d1}, but with a delayed ionization pulse. For comparison, the red ionization spectrum in the uppermost plot is taken from the simulation in Fig. \ref{fig:2d1}}. Delaying the ionization step removes the power broadening.}
        \label{fig:2d2}
    \end{subfigure}
    \caption[]
    {Two simulations for a two-level atom, using the parameters in table \ref{tab:parameters}. For each of the figures, the central surface plot shows the excited state population as a function of laser detuning and time. The diagram to the left of these central plot schematically displays the laser pulse sequence. On the top of each figure the resonance ionization spectrum is shown.} 
    \label{fig:2ds}
\end{figure*}

In a two-step RIS scheme, both lasers can broaden the resonance line profiles. Power broadening due to the excitation laser in a closed two-level system using continuous-wave (CW) laser light is well understood in the steady-state limit. In this case, the linewidth of the optical resonance increases with the laser power \cite{Citron_1977}:
\begin{equation}
    \text{FWHM} = A\sqrt{ 1+2(\Omega/A)^2 }.
\end{equation}
In other words, population is only excited efficiently to the excited state when
\begin{equation}
    \left|\Delta\right| \lesssim A\sqrt{ 1+2(\Omega/A)^2 }.\label{eq:power_broadening}
\end{equation}
However, for pulsed laser excitation, this relationship is not always valid \cite{Vitanov2001}.  It can be derived that for a Gaussian shaped excitation laser pulse (in absence of ionization and spontaneous decay), the detuning range that results in excited state population after the action of the excitation pulse is given by \cite{Boradjiev2013}
\begin{equation}
    \left|\Delta\right| \lesssim \frac{\sqrt{\log{(\Omega/\Delta)}}}{T},\label{eq:Gaussbroadening}
\end{equation}
with T the length of the laser pulse in time. The different power dependence of the line widths for CW and pulsed laser excitation is illustrated in Fig. \ref{fig:pb_eqs}, using $A=1$\,MHz in \eqref{eq:power_broadening} and $T=50$\,ns in \eqref{eq:Gaussbroadening}. The linewidth predicted by equation \eqref{eq:Gaussbroadening} for pulsed laser excitation scales much more favorably with the laser power. Since this reduced linewidth is only obtained after the excitation pulse has passed (as illustrated further in Fig. \ref{fig:2ds}), the significantly reduced power broadening presented in Fig. \ref{fig:pb_eqs} can only be obtained by using a subsequently delayed ionization pulse. The considerable reduction of the resonance linewidth provides a strong argument in favor of using an ionization step that is delayed with respect to the pulsed excitation laser step, such that the narrower line shape is probed.

Besides power broadening due to the excitation laser, the interaction of the system with the ionization laser, if applied during the excitation laser pulse, can also further broaden the level. This can be understood in an intuitive way. Since the ionization laser couples the excited state to the continuum, the lifetime of this excited state is reduced. By virtue of the Heisenberg energy-time uncertainty principle, this implies an increase in the energy uncertainty of the excited state. Indeed, the resonant excitation is probed by looking at the ions that are created by subsequent excitation of the atoms from the excited level towards the ionization continuum. The energy uncertainty induced in the intermediate level translates into a broadening of the resonance observed in the spectrum. If the ionization laser is delayed with respect to the excitation laser step, this broadening does not occur, since the perturbing ionizing laser field is not present when the resonant excitation happens.

Fig. \ref{fig:2ds} illustrates these observations by presenting numerical solutions to the equations of motion for a two-level system using the parameters presented in table \ref{tab:parameters}. The population of the excited state as a function of the laser detuning (x-axis) and time (y-axis) is shown. To the left of these plots a schematic picture of the time sequence of the excitation laser pulse and the second laser pulse is shown. The spectrum above the two-dimensional plots shows the frequency dependence of the ionization efficiency obtained at the end of the pulsed resonance ionization process. 

\begin{table}[ht!]
    \begin{tabular}{lr}
    Parameter & Value\\
    \hline
    Excitation Laser power & 10\,nJ \\
    Excitation Laser pulse length & 50\,ns \\
    Ionization Laser power & 1\,mJ \\
    Ionization Laser pulse length & 8\,ns \\
    A-parameter & 10\,MHz \\
    Photo-ionization cross section $\sigma$ & 1\,Mb\\
    Ionization laser delay for Fig. \ref{fig:2d1}, \ref{fig:2d3} & 0\,ns\\
    Ionization laser delay for Fig. \ref{fig:2d2}, \ref{fig:2d4} & 80\,ns \\
    Stark effect Fig. \ref{fig:2ds} & $S=0$\\
    Stark effect Fig. \ref{fig:2ds2} & $S=0.9\cdot\Gamma(t)$ \\
    Fano parameter q & 0 \\
    \end{tabular}
    \caption{Parameters used for the simulations in figures \ref{fig:2ds} and \ref{fig:2ds2}.}\label{tab:parameters}
\end{table}

In Fig. \ref{fig:2d1} it is shown that population is transfered to the excited state as the excitation laser builds up. When the ionization laser fires, the accumulated population is removed from the excited state. Upon comparing the ionization spectra shown at the top of figures \ref{fig:2d1} and \ref{fig:2d2}, it becomes apparent that delaying the ionization laser until the excitation laser has ended considerably reduces the linewidth of the final optical resonance obtained through the resonance ionization. This is due to the transient nature of the population transfer outside of the narrow region governed by equation \eqref{eq:Gaussbroadening}: only in that region will population remain in the excited state after the action of the excitation laser. The width of the resonance ionization signal will therefore be narrower when using a delayed ionization laser. Note also in Fig. \ref{fig:2d1} how the presence of the ionization laser leads to additional broadening of the excitation spectrum, an additional source of line broadening that is avoided by using a delayed ionization stage. 

The goal of the experiments that are described below is to illustrate how power broadening can be mitigated by using a delayed ionization step. In this demonstration, the use of a weak transition to a long-lived excited state is crucial, since spontaneous decay from the excited state is minimal even with delayed ionization. This means that the resonance ionization process still occurs efficiently, which is critical for applications on exotic radioactive beams.  

\subsection{Experimental verification}

\subsubsection{Description of the experiment}

\begin{figure}[ht!]
\begin{center}
\includegraphics[width=\columnwidth]{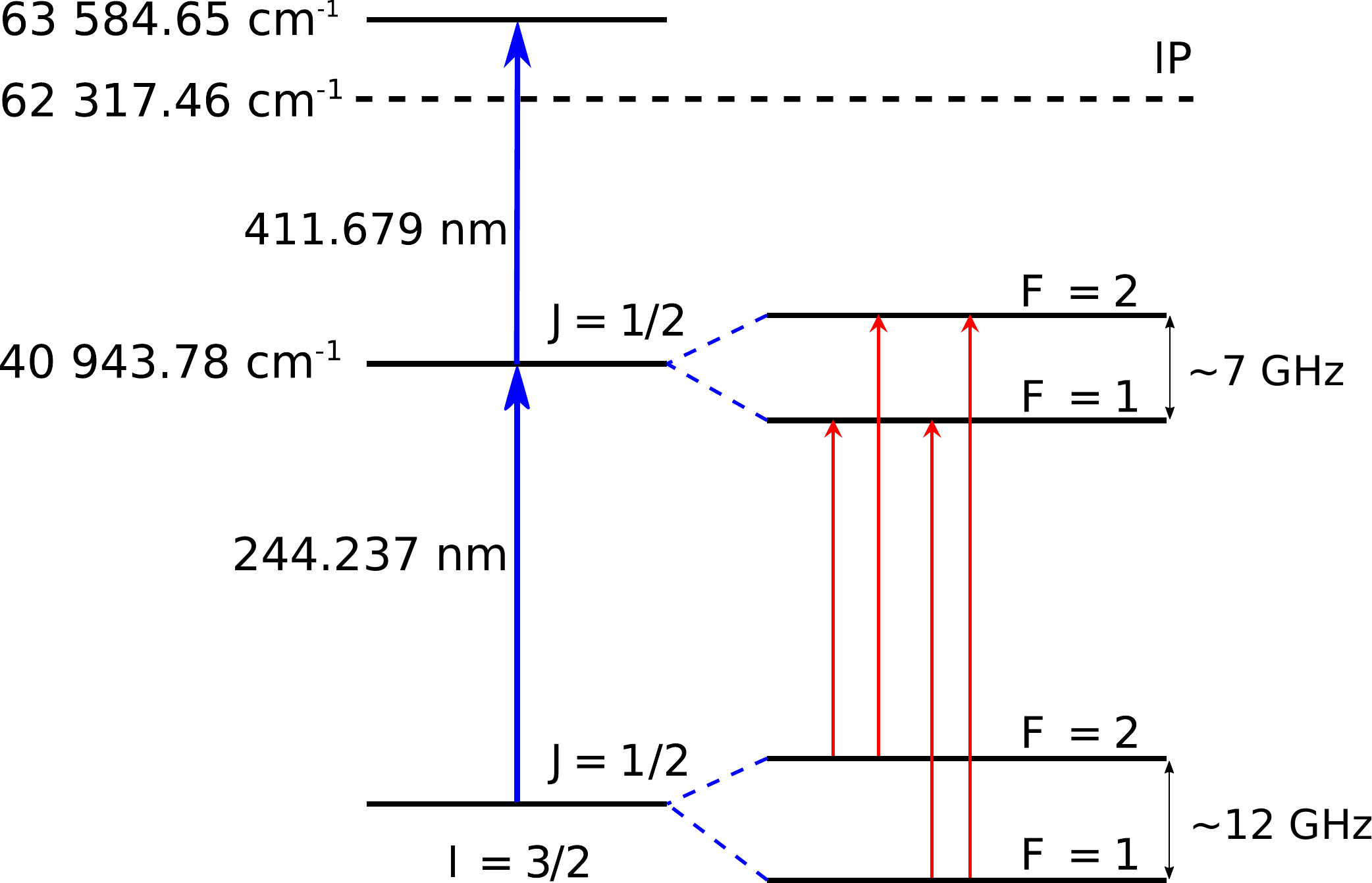}
\caption{Ionization scheme used for $^{63}$Cu and $^{65}$Cu.%
}
\label{fig:cu_scheme}
\end{center}
\end{figure}

$^{63,65}$Cu atoms have been laser-ionized using a two-step resonance ionization process depicted in Fig. \ref{fig:cu_scheme}. The experiment was performed at the JYFL laboratory in Jyv\"askyl\"a, Finland. The resonant 244.237\,nm line from the $3d^{10}4s \ ^2S_{1/2}$ ground-state to the $3d^{9}4s4p \ ^4P^o_{1/2}$ state at 40943.78\,cm$^{-1}$ was followed by a 441.679\,nm transition to the auto-ionizing $3d^{9}4s \ 5s \ ^4D_{3/2}$ state at 63584.65\,cm$^{-1}$. Given the long lifetime of the excited bound state (479(28)\,ns \cite{Kono1982}), this system is well suited to study the behavior of power broadening for pulsed lasers and the role of the delay of the ionization laser on the line shape and ionization efficiency. Furthermore, the laser system used to excite the transition had a narrow bandwidth ($\approx 20$\,MHz) and could deliver an order of magnitude more laser power than the required saturation power density, resulting in clear power broadening effects. 

A description of the atomic beam unit used for this work is presented in \cite{Kessler2008}, but the essential features of the device are repeated here. The copper atoms were produced by resistively heating a tantalum tube containing a sample of copper. The resulting atomic beam passed through a collimation slit and then orthogonally crossed the laser beams. Electrostatic ion optics extracted the laser ionized copper atoms from the interaction region and guided them to an electron multiplier tube which served as the particle detector. 

\begin{figure}[ht!]
\begin{center}
\includegraphics[width=\columnwidth]{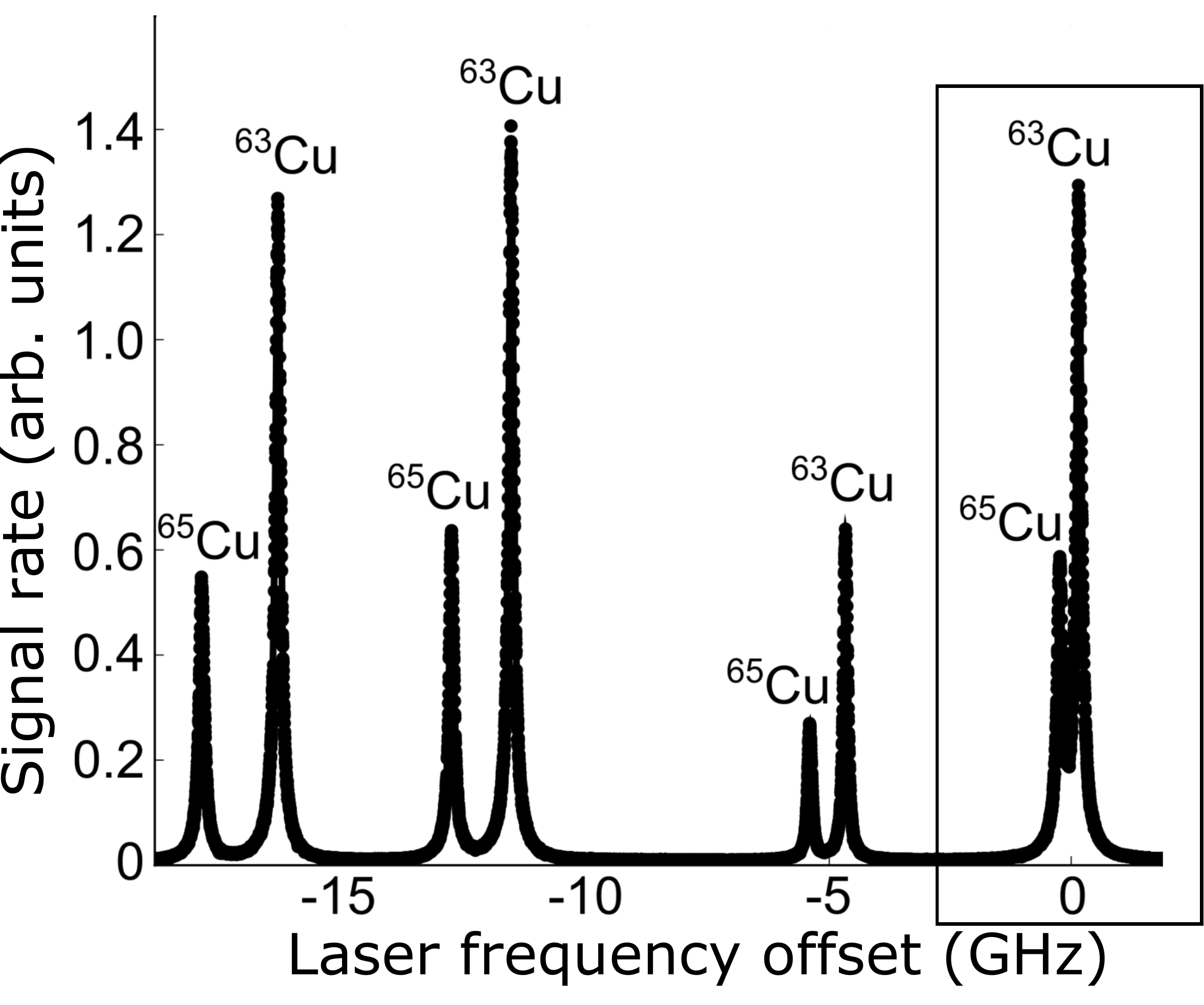}
\caption{Ionization spectrum of $^{63,65}$Cu. The black box indicates the peaks presented in the zoom-in in Fig. \ref{fig:comparison}.}
\label{fig:full_scan}
\end{center}
\end{figure}

The laser system used for this work is described in detail in \cite{Sonnenschein2015}. For the 244.237\,nm line, an injection-locked Ti:sapphire laser system produced narrowband laser light (bandwidth $\approx$ 20\,MHz), which was then frequency tripled. The master laser for this seeding cavity was a CW Matisse Ti:sapphire laser, which can be scanned continuously. The fundamental output of the seeded laser was 2.8\,W at 10\,kHz repetition rate, which after beam transport losses resulted in 300\,mW/cm$^2$ of tripled UV light entering into the atomic beam unit. A maximum of 1.6\,W/cm$^2$ of 411.679\,nm light for the ionization step was produced using an intra-cavity frequency doubled pulsed Ti:sapphire laser. The two lasers were pumped using different Nd:YAG lasers, which introduces a jitter in the timing synchronization of both Ti:sapphire lasers of about 10\,ns. This time jitter was of no consequence for the experiment. The pulse length of both lasers is typically 50\,ns. The wavelength of the injection seeded Ti:sapphire was recorded using a High Finesse WS6 wavemeter and further monitored with a Toptica scanning Fabry Perot Interferometer FPI-100-0750-y with a free spectral range of 1\,GHz. This interferometer was used to more precisely determine the wavelength of the laser as it was scanned. An example of a resonance ionization spectrum of $^{63,65}$Cu is shown in Fig. \ref{fig:full_scan},

\subsubsection{Discussion of results}

\begin{figure}[ht!]
\begin{center}
\includegraphics[width=\columnwidth]{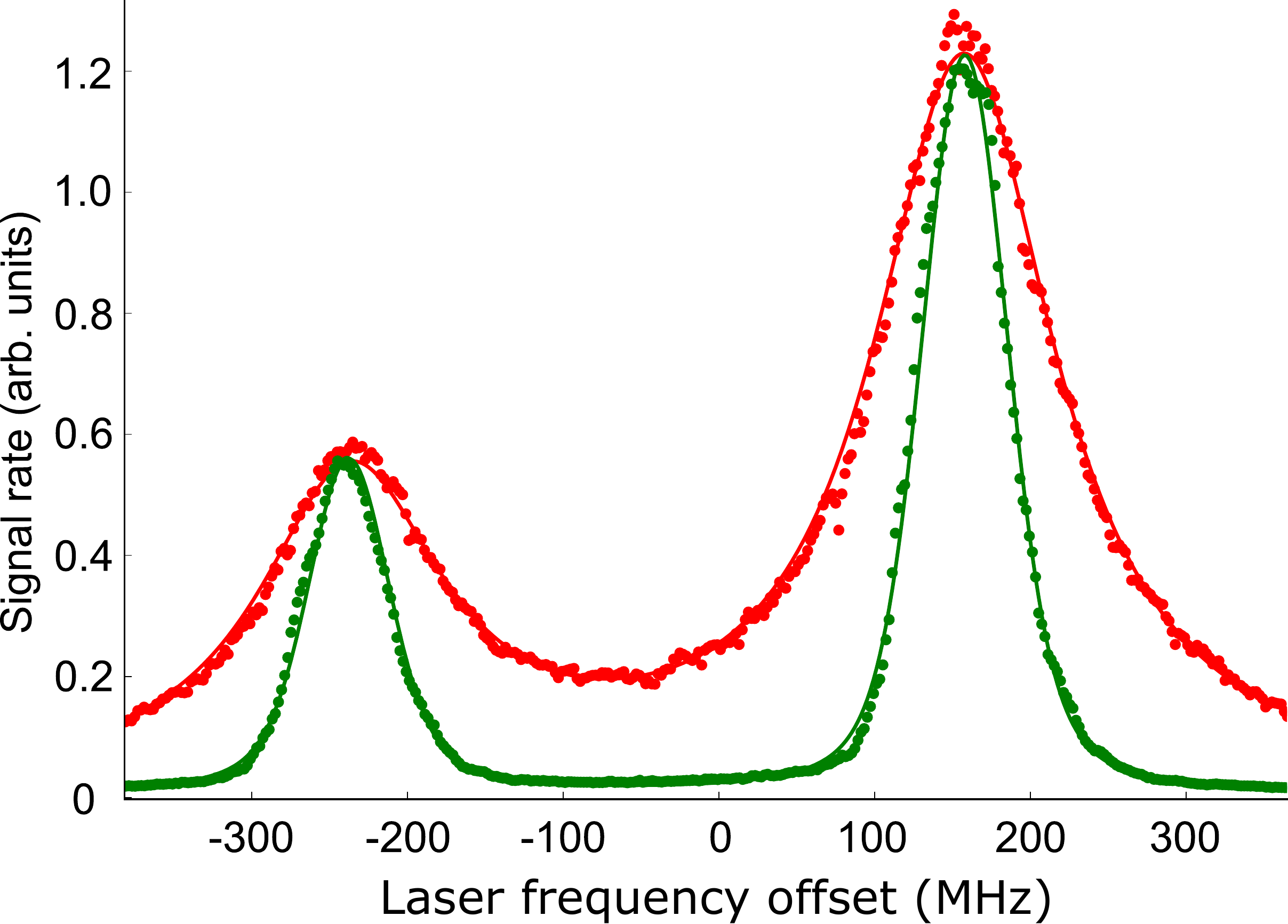}
\caption{Two lines in the spectrum of $^{63,65}$Cu (see Fig. \ref{fig:full_scan} for the full spectrum). Plotted in red is a spectrum obtained with temporally overlapping lasers (broad lines), while green is used for a spectrum obtained with a delayed ionization step (40(10)\,ns delay). Symbols are the experimental data points, the full lines are the fit. The laser power was 150 mW / cm$^2$ for both measurements. A sharp reduction in the linewidth can be clearly seen, without loss in efficiency.
}
\label{fig:comparison}
\end{center}
\end{figure}

Ionization spectra of $^{63,65}$Cu were obtained at different UV laser powers and for several time delays of the ionization step  with respect to the excitation step. The linewidth of the Gaussian component of the fitted Voigt profiles was found to vary between 40 and 60\,MHz for all experimental conditions. The contributions to this Gaussian component from the remaining Doppler broadening and the laser linewidth could not be separated, but are likely of a similar magnitude. The measurement performed at the lowest UV laser power (3\,mW/cm$^2$) and using a temporally overlapped excitation and ionization laser resulted in a Lorentzian component with a width of 53.8(4)\,MHz. Increasing the power of the laser to 150\,mW/cm$^2$ increased the linewidth of the Lorentzian component to 124.3(3)\,MHz. The two right-most hyperfine transitions in the spectrum of $^{63,65}$Cu measured with this larger laser power are shown in Fig. \ref{fig:comparison}. The spectrum in red is measured with the excitation and ionization lasers firing simultaneously, while the green spectrum was obtained by delaying the ionization laser 40(10)\,ns. Symbols are the experimental data points, the full line is the fit with a Voigt line shape. As can be clearly seen by comparing the two spectra, keeping the laser power fixed and delaying the ionization laser drastically reduces the width of the resonance lines.

\begin{figure}[ht!]
\begin{center}
\includegraphics[width=\columnwidth]{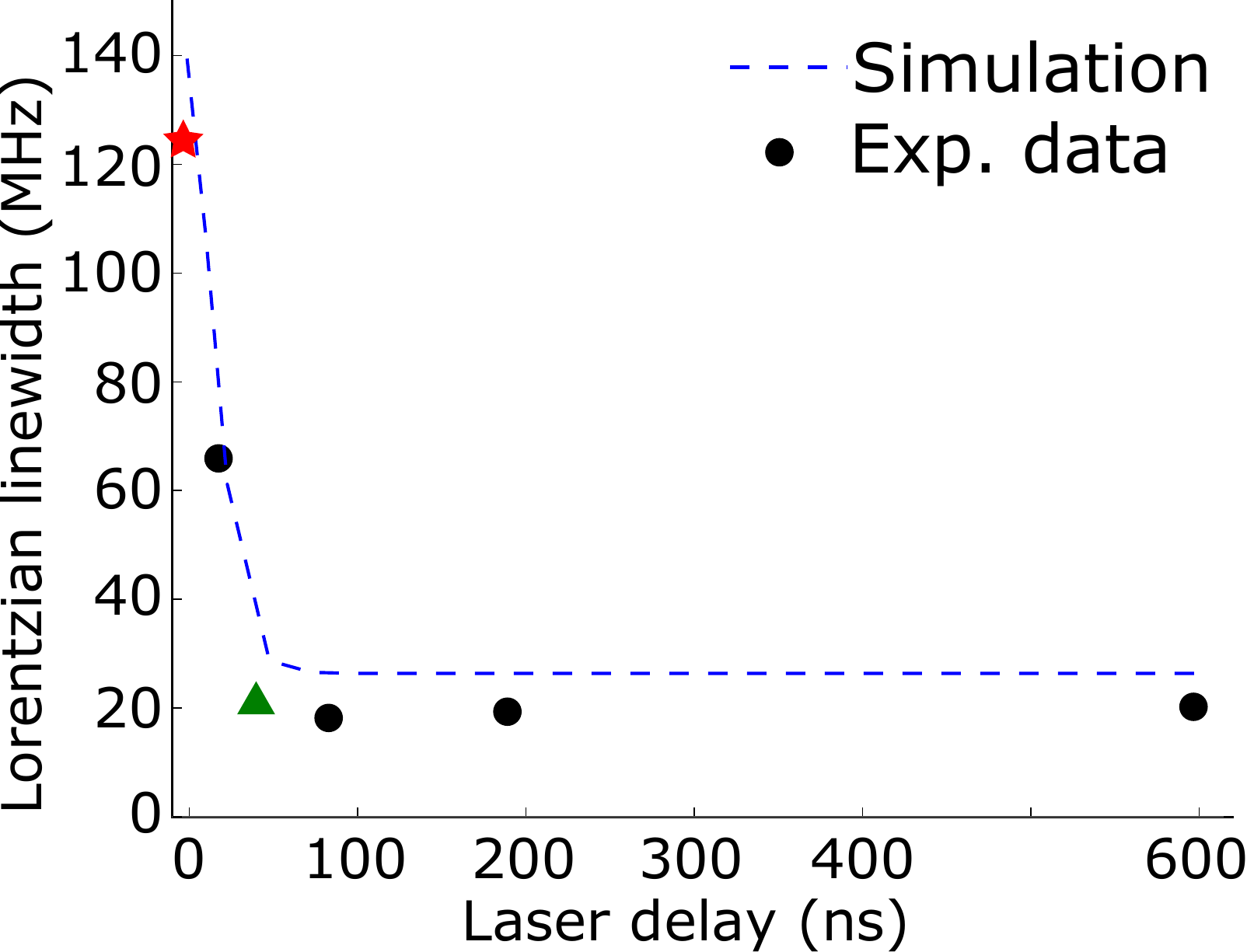}
\caption{Linewidth of the Lorentzian component of the Voigt profile as a function of the delay time of the second laser pulse. The red star corresponds to the red spectrum in Fig. \ref{fig:comparison}, the green triangle corresponds to the green spectrum in Fig. \ref{fig:comparison}. Errors are smaller than the symbols. Also shown is a theoretical calculation of the linewidth as function of the laser delay, using the experimental parameters described in the text.
}
\label{fig:graph}
\end{center}
\end{figure}

This sharp reduction in the experimental linewidth as the laser is delayed is further illustrated in Fig. \ref{fig:graph}. This figure shows the linewidth of the Lorentzian component as a function of the delay of the second laser. The red star and green triangle in this graph correspond to the data plotted in red and green of the spectra in Fig. \ref{fig:comparison}. Also shown on Fig. \ref{fig:graph} is a theoretical simulation, using the experimental parameters given earlier in this text. Even though there is an offset between theory and experiment, the general trend is well reproduced, indicating that the cause of the reduction in linewidth is understood. The experimental Lorentzian linewidth saturates at about 19\,MHz for delay times of more than 100\,ns. Note that this reduced Lorentzian linewidth is much less than the 53.8(4)\,MHz linewidth obtained at the lowest laser power of 3\,mW/cm$^2$ (with temporally overlapping laser pulses). This provides direct evidence for the absence of power broadening not only from the excitation laser but also from the ionization laser. 

Note finally that the efficiency loss due to spontaneous decay is negligible for delays below 50\,ns due to the long lifetime of the excited state of 479(28)\,ns. This is the key advantage offered by using the weak transition to a long-lived state rather than a stronger transition.

\section{Experiments on lineshape distortions and delayed ionization}\label{sec:dist}

\subsection{Model Predictions}

\begin{figure*}[ht!]
    \centering
    \begin{subfigure}[b]{0.485\textwidth}   
        \centering 
        \includegraphics[width=\textwidth]{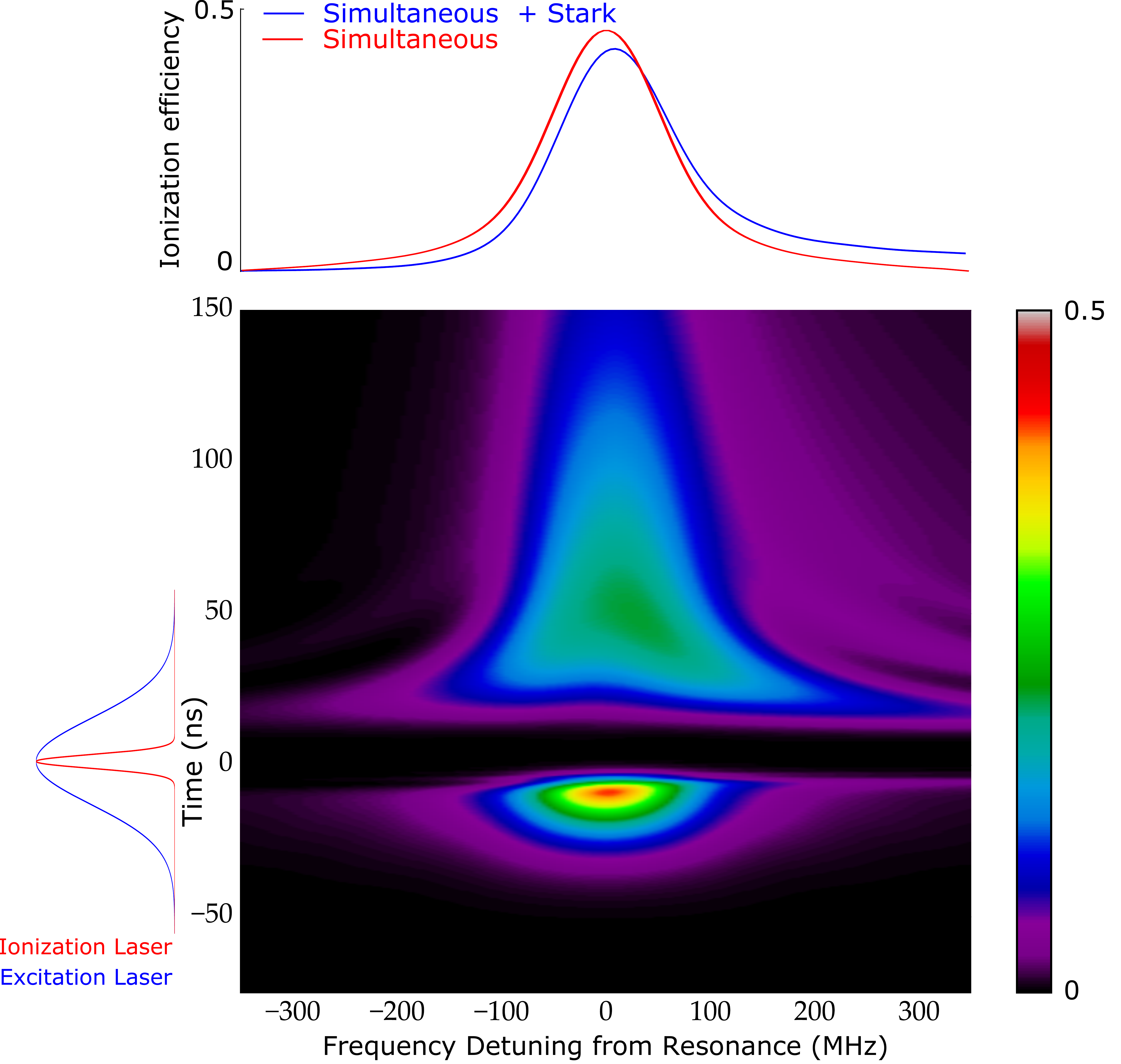}
        \caption[]%
        {{Excited state population and ionization spectrum for simultaneous laser pulses. For comparison, the red curve in the top plot is taken from figure \ref{fig:2d1}}. This comparison shows a clear asymmetry in the ionization spectrum (shown as the blue curve in the top graph), which can also be seen in the population of the excited state just after the second laser pulse.}
        \label{fig:2d3}
    \end{subfigure}
    \quad
    \begin{subfigure}[b]{0.485\textwidth}   
        \centering 
        \includegraphics[width=\textwidth]{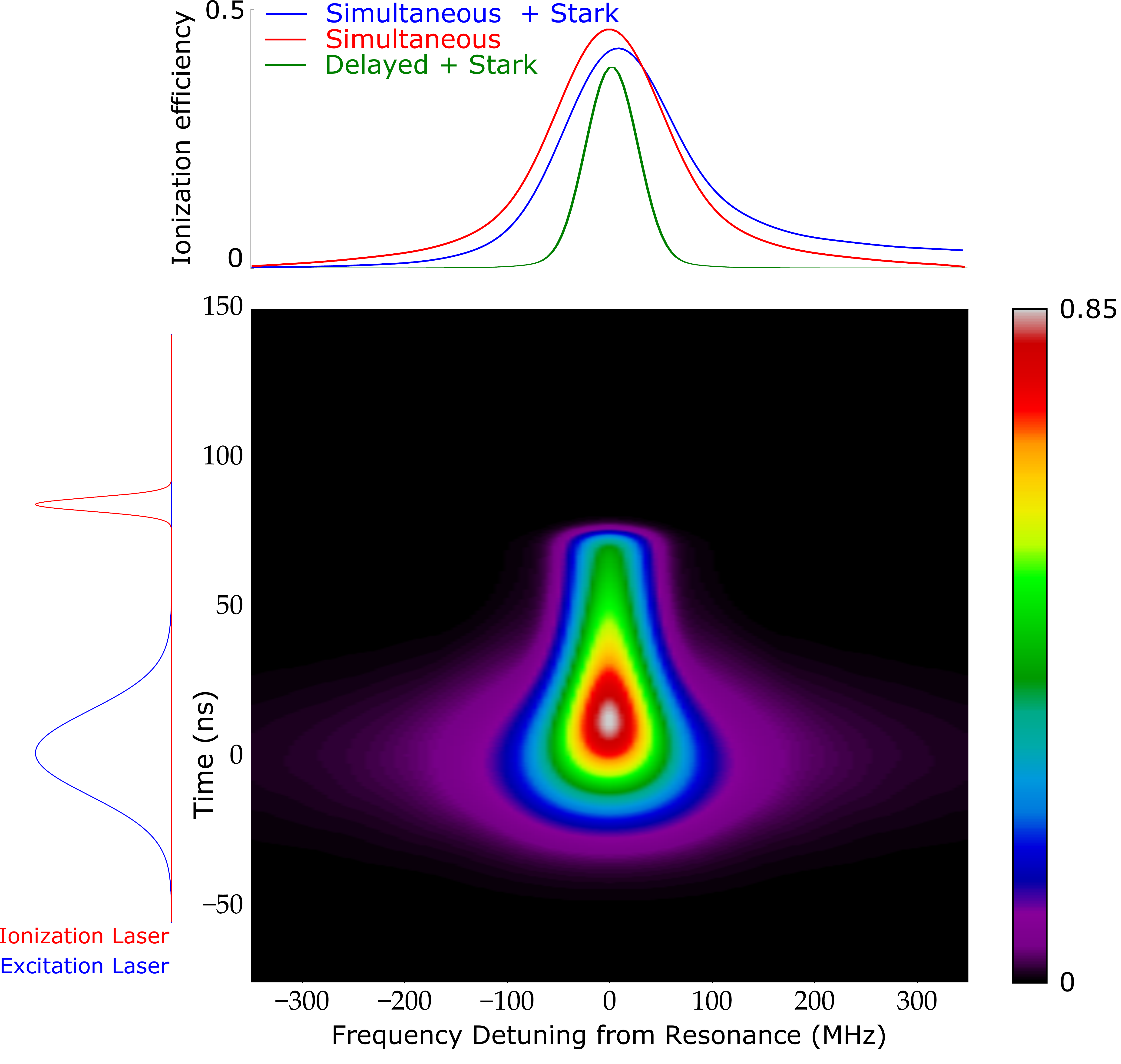}
        \caption[]%
        {{Excited state population and ionization spectrum for simultaneous laser pulses. For comparison, he red and green curves in the top plot are taken from figures \ref{fig:2d1} and \ref{fig:2d2} respectively. By delaying the ionization laser, both the power broadening and asymmetry in the ionization spectrum disappear.}}
        \label{fig:2d4}
    \end{subfigure}
    \caption[]
    {As figure \ref{fig:2ds}, but including a Stark shift due to the ionization laser. } 
    \label{fig:2ds2}
\end{figure*}

In addition to the power broadening effects discussed in the previous section, there is another effect to consider: the possibility of lineshape distortions induced by a high-power ionization laser. This effect is illustrated in Fig. \ref{fig:2ds2}. This figure repeats the simulations presented earlier in Fig. \ref{fig:2ds}, but this time includes a Stark shift induced by the ionization laser ($S$ in equation \eqref{eq:ionization_ham}). In Fig. \ref{fig:2d3} a clear asymmetry can be seen in both the population of the excited state and the final ionization spectrum. By contrast, when using a delayed ionization laser, this asymmetry is naturally absent. 

The next section will discuss some experimental data which demonstrate this kind of lineshape distortions. We will also show that the model for laser ionization introduced in section \ref{sec:intro} can be used to qualitatively explain these distortions. Delaying the ionization laser in time with respect to the excitation laser removes the unwanted effects, since the distortion is induced after the atomic structure is already probed by the excitation laser. 

\subsection{Experimental verification}

\subsubsection{Description of the experiment}

\begin{figure}[ht!]
\begin{center}
\includegraphics[width=\columnwidth]{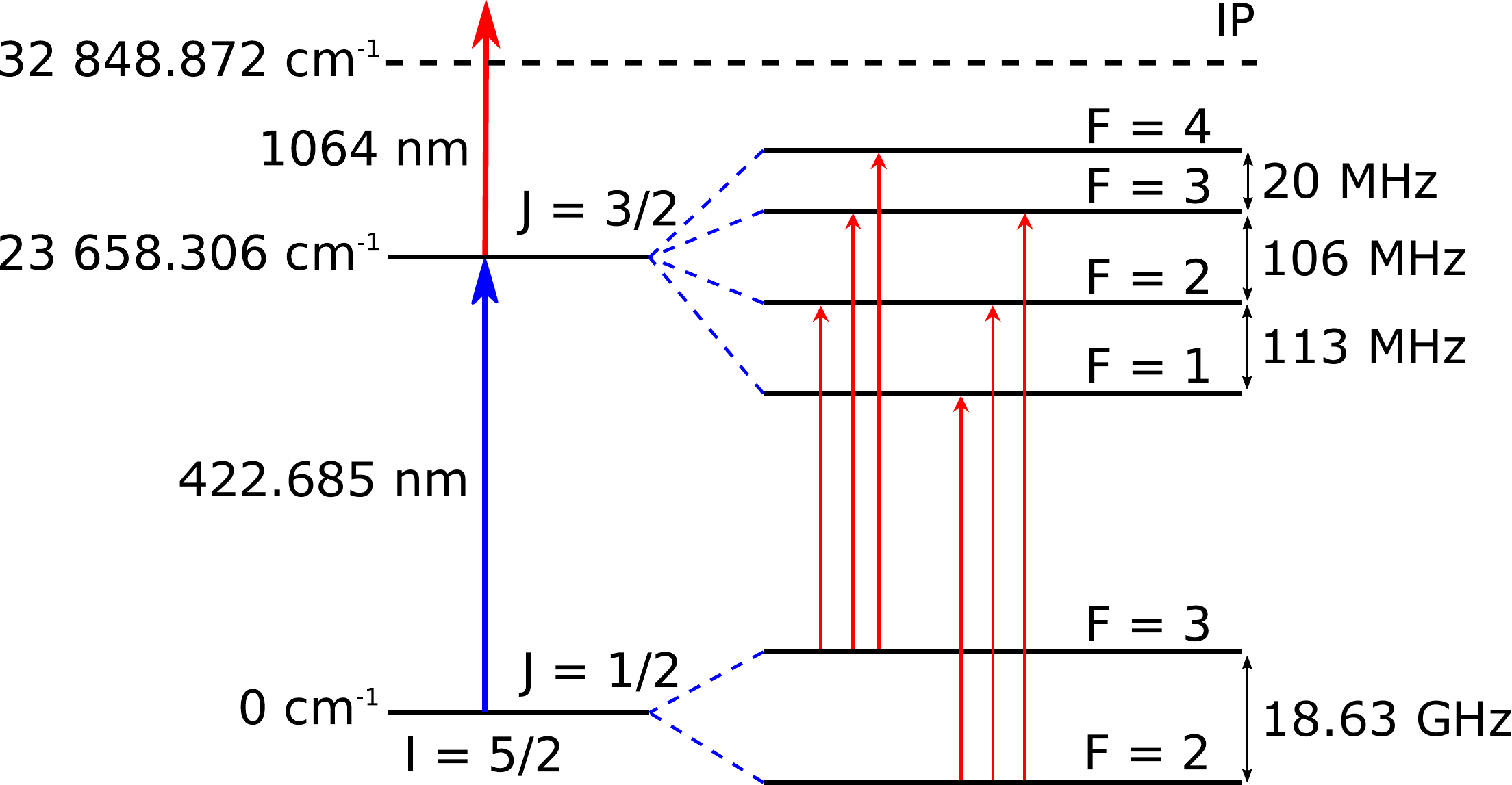}
\caption{Ionization scheme used to excite and ionize $^{221}$Fr. The size of the hyperfine splittings was calculated using the values in \cite{Duong1987}. %
}
\label{fig:fr_scheme}
\end{center}
\end{figure}

The possibility of ionization-related lineshape distortions, and how they can be removed by using a delayed ionization step, has been illustrated using the Collinear Resonance Ionization Spectroscopy (CRIS) experiment at ISOLDE-CERN \cite{de_Groote_2015,Flanagan2013}, using a radioactive beam of $^{221}$Fr.  Details on the experimental set-up and measurement procedure can be found in \cite{de_Groote_2015}. The ionization scheme that was used is presented in Fig. \ref{fig:fr_scheme}, and consists of an excitation step from the $7s\ {}^2S_{1/2}$ ground state to the $8p\ {}^2 P_{3/2}$ state at 23658.306\,cm$^{-1}$ (422.685\,nm), and an ionization step that non-resonantly ionizes from the $8p\ {}^2 P_{3/2}$ state using pulsed 1064\,nm light. The lifetime of the excited atomic state is 83.5(1.5)\,ns, sufficiently long to justify the use of a delayed ionization pulse.

A $^{221}$Fr ion beam was produced by the ISOLDE facility at CERN by impinging 1.4\,GeV protons onto a uranium carbide target. Francium atoms diffuse out of this target and are then surface ionized in a hot capillary tube. After mass separation from the other francium isotopes, the $^{221}$Fr ion beam is then guided to a gas-filled linear Paul trap, where it is cooled and bunched. This beam is then accelerated to 30\,kV and transported towards the CRIS experiment. The  first stage of the CRIS experiment consists of neutralizing the ion beam through collisionless charge exchange with a hot potassium vapor \cite{Procter2012}. This is often required, since suitable transitions are usually easier to find for neutral atoms rather than ions. The non-neutralized fraction of the beam is electrostatically deflected into a beam dump while the neutralized fraction is temporally and spatially overlapped with the laser beams in an ultra-high vacuum (UHV) interaction region. The laser frequency of the UV excitation laser is scanned across the hyperfine resonance transitions, and the resonantly ionized $^{221}$Fr ions are then deflected onto a copper dynode. The secondary electrons emitted from the dynode are detected using a microchannel plate (MCP) electron detector. The UHV is required to minimize collisional ionization that would otherwise result in a constant background in the hyperfine spectra. 

Because of the combination of an accelerated beam and the collinear overlap of the atom and laser beams, Doppler broadening is reduced to the point where it only contributes a few MHz to the total linewidth of the hyperfine structure spectra. The laser light for the first step was produced by frequency doubling the output of a Matisse TS cw Ti-sapphire laser with a Wavetrain external cavity frequency doubler. This continuous light was chopped into pulses of variable length through the use of a pockels cell and subsequent polarization sensitive beam optics, described in detail in \cite{de_Groote_2015}. This experimental configuration was used to create light pulses with a pulse length of $100$\,ns, at a repetition rate of 100\,Hz. The 1064\,nm light for non-resonant ionization was produced using a dual-cavity Litron LPY 601 50-100 PIV laser system, operated at 100\,Hz and with a pulse length of 13\,ns. After beam transport losses, 250\,mW/cm$^2$ of continuous wave laser light and 32\,mJ/pulse of 1064\,nm laser light reached the entry of the CRIS beamline.

\subsubsection{Discussion of results}

\begin{figure}[ht!]
\begin{center}
\includegraphics[width=\columnwidth]{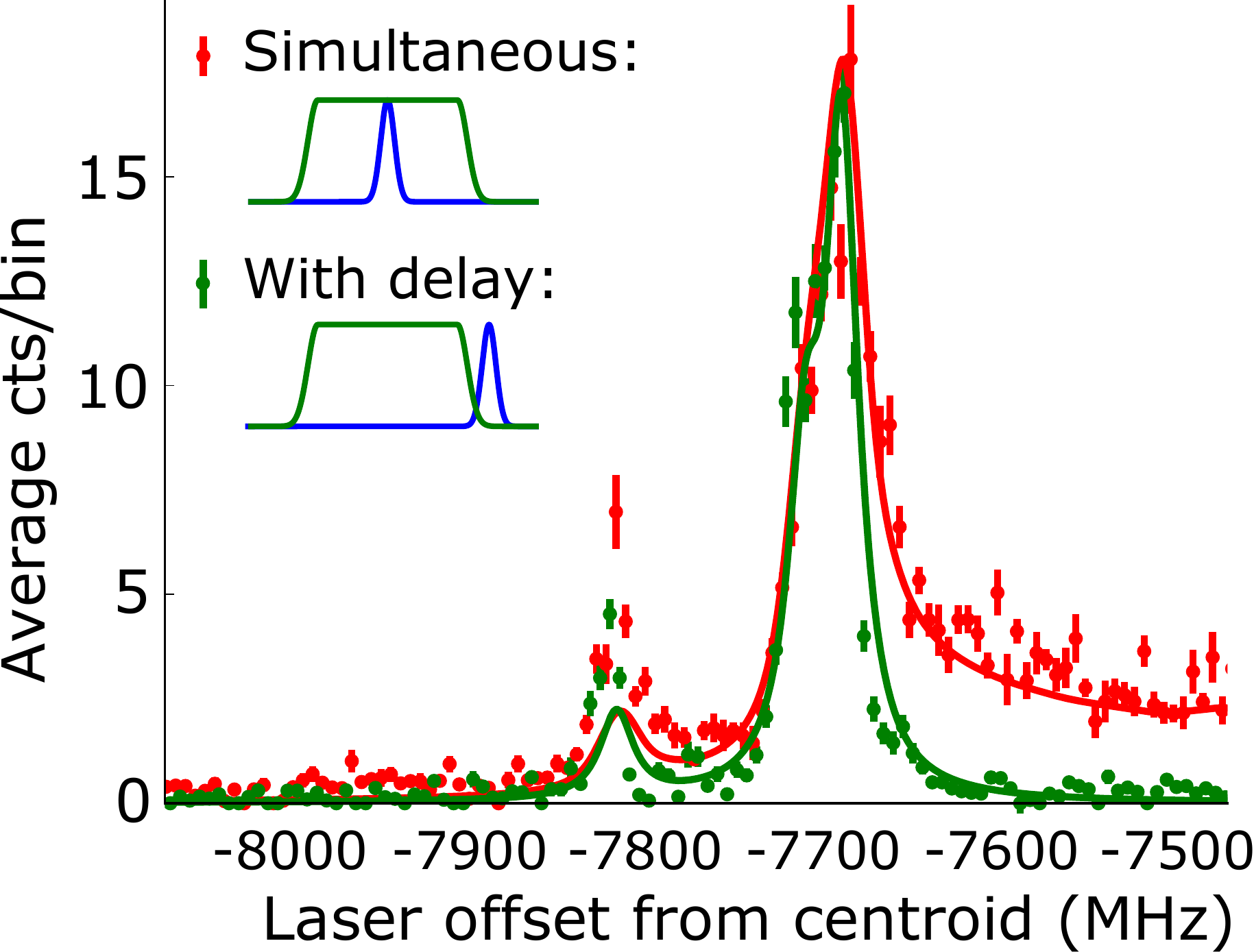}
\caption{Resonance ionization spectra of the leftmost components of the hyperfine structure of $^{221}$Fr, obtained with simultaneous laser pulses (red) and with a delayed ionization step (green). The solid lines are fits using the model for RIS presented in section \ref{sec:model}. The asymmetry of the line disappears when the ionization laser is delayed, while the total ionization efficiency does not decrease significantly.}
\label{fig:fr_data}
\end{center}
\end{figure}

Fig. \ref{fig:fr_data} shows two measurements of the low-frequency component of the hyperfine structure of $^{221}$Fr. The red (broad) spectrum is obtained with the ionization laser temporally overlapped with the 100\,ns wide excitation laser pulse, as illustrated in the inset of Fig. \ref{fig:fr_data}. The green spectrum was obtained with an ionization pulse  delayed by 100\,ns from the start of the excitation pulse. Using simultaneous laser pulses distorts the high-frequency side of the peaks, which displays a clear asymmetry. This asymmetry disappears when the ionization laser is delayed, which indicates that the tailing is induced by the ionization laser.

The figure also shows simulations using the model introduced in section \ref{sec:model}. The ionization cross section $\sigma$ was taken to be 1\,Mb, which should at least be of the correct order of magnitude (see e.g. \cite{Ambartzumian1976,Gilbert1984} for cross sections in Rb and Cs). The Fano $q$ parameter was taken to be zero. The effective Stark shift $S$ was tuned to give the best match with the experimental data; a final value of $S(t) = 4 \Gamma(t)$ provided good agreement. The simulations were also rescaled to match the intensity of the highest peak in the experimental data. 

Using these parameters, the asymmetric tail of the peaks is well reproduced, supporting the idea that the observed asymmetry is due to a Stark shift caused by the strong electric field of the high-power ionization laser. The intensity of the smallest resonance in the spectrum is not well reproduced by the model. The reason for this discrepancy is unclear.

As with the data on the copper isotopes, delaying the ionization step does not result in significant loss in efficiency, since the excited state is long-lived. The linewidth of the resonance is 20(1)\,MHz. This linewidth could only be reached due to the removal of power broadening and the lineshape-distorting AC Stark shift by delaying the ionization laser. In a two-step resonance ionization scheme, this can only be done efficiently with a weak transition to a sufficiently long-lived excited state.

\section{Efficient laser excitation and ionization with weak transitions}\label{sec:eff}

Weak transitions have shown desirable features for laser spectroscopy purposes. In addition to the inherently small linewidth, weak transitions show no sign of efficiency loss when delaying the ionization pulse. This section will further argue that weak transitions to long-lived states can be excited with very high absolute efficiencies, comparable to efficiencies obtained with stronger lines. 

Applying the model of section \ref{sec:model}, one obtains the steady state population of  an excited level in a two-level approximation as:
\begin{align}
 P_{exc}(\Delta=0) &=  \dfrac{\Omega^2}{A^2+2\Omega^2}\\
& \propto \dfrac{I/A}{1+2I/A},
\end{align}\label{eq:steadystate}
since $\Omega \propto \sqrt{IA}$. Since, for a fixed laser intensity, the equilibrium population is a monotonically decreasing function of $A$, weak transitions can achieve higher steady state population in the excited state. However, the irradiation time required to reach this steady state is longer than for strong transitions, though it also decreases with laser power. Therefore, there are two strategies to consider when maximizing the efficiency of excitations using weak transitions. 

First of all, one can use high power pulsed laser systems which increase the rate at which the equilibrium population is reached, resulting in higher efficiency for short pulse lengths. This is the approach used for the first dataset in this article (see section \ref{sec:pb}): a high-power pulsed laser can saturate the excitation step and therefore efficiently excite the system.

On the other hand, employing low power chopped CW laser light with long interaction times, as in the experiment of \ref{sec:dist}, also allows for high efficiency ($>$1-10 \%). Indeed, the efficiency in the experiment described above using chopped CW laser light \cite{de_Groote_2015} is similar to that obtained in an earlier experiment using a pulsed high-power laser for the excitation step \cite{Flanagan2013}. In both experiments, the total efficiency was 1\%, where the detection efficiency was 80\%, beam transport efficiency $<$30\%, neutralization efficiency $<$50\%, and the laser ionization efficiency therefore $>8\%$. 

\begin{figure}[ht!]
\begin{center}
\includegraphics[width=\columnwidth]{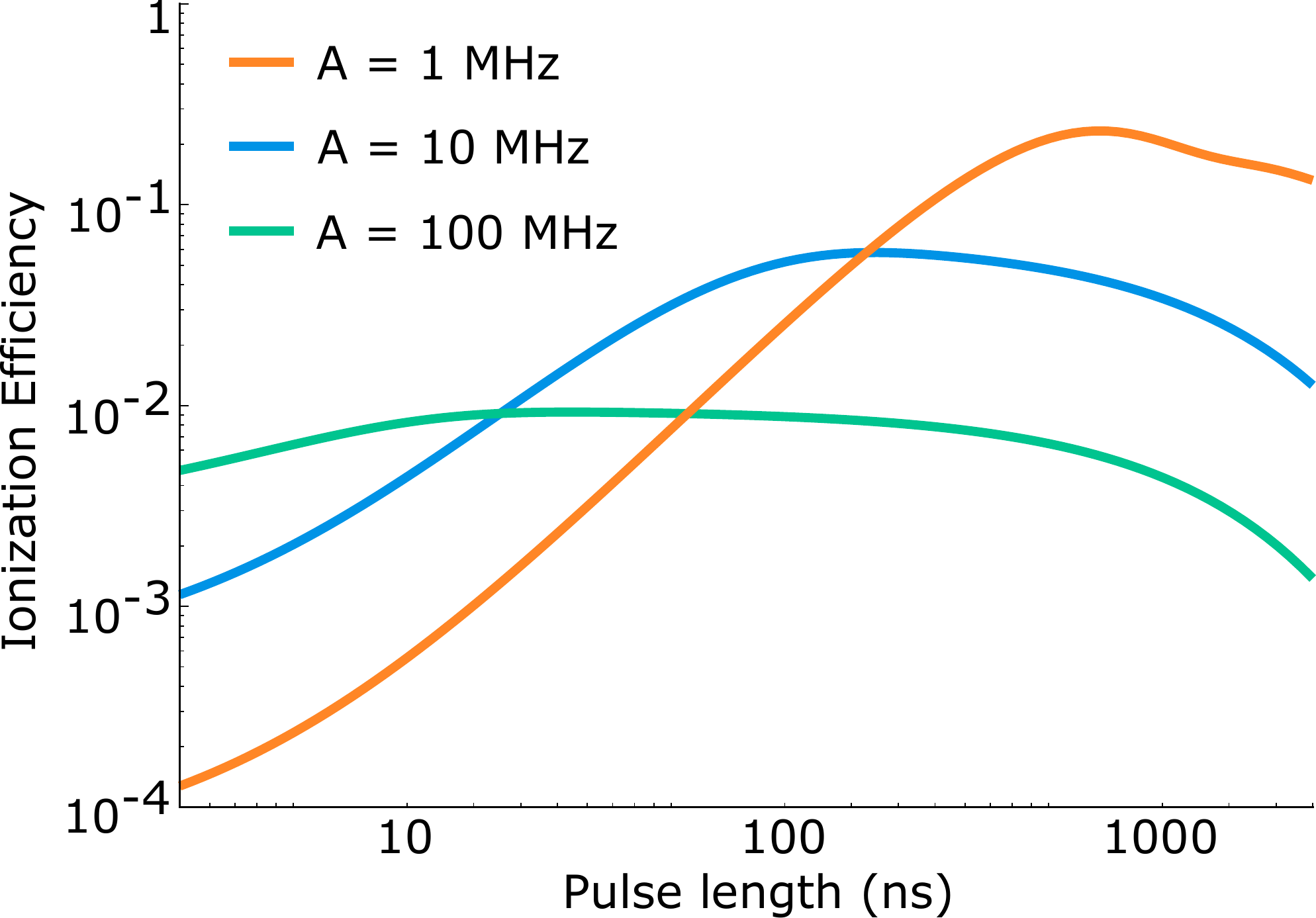}
\caption{Ionization efficiency for a chopped cw-laser with a cw power of 10\,mW, overlapped with a 1\,mJ/pulse ionization laser and with a variable excitation laser pulse length. The system that was simulated here consists of a ground state doublet and a single excited state. For a weak transition ($A=1$\,MHz) the highest efficiency is reached if the pulse length is $\approx 1\,\mu s$. For extremely long laser pulses, the efficiency decreases in all cases due to optical pumping towards the other hyperfine level of the ground state.}
\label{fig:pulselength}
\end{center}
\end{figure}

The ability to reach high laser ionization efficiencies with chopped cw laser pulses is also illustrated in Fig. \ref{fig:pulselength}. This figure shows simulated ionization efficiencies for weak and strong transitions for a system with a ground state doublet and a single excited state, as a function of the pulse length of the excitation step, and using overlapping excitation and ionization lasers. The laser power was set to 10\,mW, which is usually easily achieved with modern CW lasers. With sufficiently long interaction times, weak transitions can be excited very efficiently, with simulated ionization efficiencies better than or comparable to the efficiency obtained for short pulses and strong transitions. For time-separated laser beams, the ionization efficiency for a strong transition never reaches that of the weak transition, due to decay losses.

\section{Conclusions}

This article has shown that efficient resonance ionization spectroscopy with high resolving powers can be achieved with pulsed (or chopped) laser beams using weak transitions to long-lived states. This was demonstrated using simulations, supported by experimental observations and illustrates the twofold advantages that weak transition to long-lived states offer.

Firstly, it is possible to remove virtually all power broadening due to both lasers in a two-step RIS scheme by choosing a suitable delay between the excitation and the ionization laser pulses. Secondly, lineshape distortions due to the presence of a strong ionizing laser field when using non-resonant ionization can cause significant lineshape distortions, which can also be removed by delaying the ionization step. Similar arguments can be made in the case of RIS schemes that use more than two lasers, but this lies outside of the scope of the work presented here.

The long lifetime of the excited state ensures that no significant efficiency losses occur due to the delay of the ionization laser. Experimental evidence for both advantages was presented and compared to simulations with a theoretical model for resonance ionization spectroscopy. The experimental data also illustrate that high efficiencies can be obtained using weak transitions.

\begin{acknowledgments}
We acknowledge the support of the ISOLDE collaboration and technical teams. We are grateful to the COLLAPS collaboration for the use of their CW Ti:sapphire laser system and wavetrain doubling unit. We thank Wouter Gins for fruitful discussions and for comparisons to simulations with rate equation codes. This work was supported by the BriX Research Program No.~P7/12 and FWO-Vlaanderen (Belgium) and GOA 15/010 from KU Leuven, ERC Consolidator Grant no.~648381, the Science and Technology Facilities Council consolidated grant ST/F012071/1 and continuation grant ST/J000159/1, and the EU Seventh Framework through ENSAR(506065). K.~T.~F. was supported by STFC Advanced Fellowship Scheme Grant No.~ST/G006415/1. This work was also supported by the Academy of Finland under the Center of Excellence Programme 20122017 (Nuclear and Accelerator Based Physics Research at JYFL).
\end{acknowledgments}

\bibliography{biblio}

\begin{thebibliography}{27}%
\makeatletter
\providecommand \@ifxundefined [1]{%
 \@ifx{#1\undefined}
}%
\providecommand \@ifnum [1]{%
 \ifnum #1\expandafter \@firstoftwo
 \else \expandafter \@secondoftwo
 \fi
}%
\providecommand \@ifx [1]{%
 \ifx #1\expandafter \@firstoftwo
 \else \expandafter \@secondoftwo
 \fi
}%
\providecommand \natexlab [1]{#1}%
\providecommand \enquote  [1]{``#1''}%
\providecommand \bibnamefont  [1]{#1}%
\providecommand \bibfnamefont [1]{#1}%
\providecommand \citenamefont [1]{#1}%
\providecommand \href@noop [0]{\@secondoftwo}%
\providecommand \href [0]{\begingroup \@sanitize@url \@href}%
\providecommand \@href[1]{\@@startlink{#1}\@@href}%
\providecommand \@@href[1]{\endgroup#1\@@endlink}%
\providecommand \@sanitize@url [0]{\catcode `\\12\catcode `\$12\catcode
  `\&12\catcode `\#12\catcode `\^12\catcode `\_12\catcode `\%12\relax}%
\providecommand \@@startlink[1]{}%
\providecommand \@@endlink[0]{}%
\providecommand \url  [0]{\begingroup\@sanitize@url \@url }%
\providecommand \@url [1]{\endgroup\@href {#1}{\urlprefix }}%
\providecommand \urlprefix  [0]{URL }%
\providecommand \Eprint [0]{\href }%
\providecommand \doibase [0]{http://dx.doi.org/}%
\providecommand \selectlanguage [0]{\@gobble}%
\providecommand \bibinfo  [0]{\@secondoftwo}%
\providecommand \bibfield  [0]{\@secondoftwo}%
\providecommand \translation [1]{[#1]}%
\providecommand \BibitemOpen [0]{}%
\providecommand \bibitemStop [0]{}%
\providecommand \bibitemNoStop [0]{.\EOS\space}%
\providecommand \EOS [0]{\spacefactor3000\relax}%
\providecommand \BibitemShut  [1]{\csname bibitem#1\endcsname}%
\let\auto@bib@innerbib\@empty
\bibitem [{\citenamefont {Bracco}\ \emph {et~al.}(2010)\citenamefont {Bracco},
  \citenamefont {Chomaz}, \citenamefont {Gaardh{{\o{}}}je}, \citenamefont
  {Makarow}, \citenamefont {Heenen}, \citenamefont {Rosner}, \citenamefont
  {Kaiser}, \citenamefont {MacGregor}, \citenamefont {Widmann},\ and\
  \citenamefont {Korner}}]{nupecc2010}%
  \BibitemOpen
  \bibfield  {author} {\bibinfo {author} {\bibfnamefont {A.}~\bibnamefont
  {Bracco}}, \bibinfo {author} {\bibfnamefont {P.}~\bibnamefont {Chomaz}},
  \bibinfo {author} {\bibfnamefont {J.}~\bibnamefont {Gaardh{{\o{}}}je}},
  \bibinfo {author} {\bibfnamefont {M.}~\bibnamefont {Makarow}}, \bibinfo
  {author} {\bibfnamefont {P.-H.}\ \bibnamefont {Heenen}}, \bibinfo {author}
  {\bibfnamefont {G.}~\bibnamefont {Rosner}}, \bibinfo {author} {\bibfnamefont
  {R.}~\bibnamefont {Kaiser}}, \bibinfo {author} {\bibfnamefont
  {D.}~\bibnamefont {MacGregor}}, \bibinfo {author} {\bibfnamefont
  {E.}~\bibnamefont {Widmann}}, \ and\ \bibinfo {author} {\bibfnamefont
  {G.}~\bibnamefont {Korner}},\ }\href@noop {} {\  (\bibinfo {year}
  {2010})}\BibitemShut {NoStop}%
\bibitem [{\citenamefont {Ruiz}\ \emph {et~al.}(2016)\citenamefont {Ruiz},
  \citenamefont {Bissell}, \citenamefont {Blaum}, \citenamefont {Ekstr{\"o}m},
  \citenamefont {Fr{\"o}mmgen}, \citenamefont {Hagen}, \citenamefont {Hammen},
  \citenamefont {Hebeler}, \citenamefont {Holt}, \citenamefont {Jansen} \emph
  {et~al.}}]{ruiz2016}%
  \BibitemOpen
  \bibfield  {author} {\bibinfo {author} {\bibfnamefont {R.~G.}\ \bibnamefont
  {Ruiz}}, \bibinfo {author} {\bibfnamefont {M.}~\bibnamefont {Bissell}},
  \bibinfo {author} {\bibfnamefont {K.}~\bibnamefont {Blaum}}, \bibinfo
  {author} {\bibfnamefont {A.}~\bibnamefont {Ekstr{\"o}m}}, \bibinfo {author}
  {\bibfnamefont {N.}~\bibnamefont {Fr{\"o}mmgen}}, \bibinfo {author}
  {\bibfnamefont {G.}~\bibnamefont {Hagen}}, \bibinfo {author} {\bibfnamefont
  {M.}~\bibnamefont {Hammen}}, \bibinfo {author} {\bibfnamefont
  {K.}~\bibnamefont {Hebeler}}, \bibinfo {author} {\bibfnamefont
  {J.}~\bibnamefont {Holt}}, \bibinfo {author} {\bibfnamefont {G.}~\bibnamefont
  {Jansen}},  \emph {et~al.},\ }\href@noop {} {\bibfield  {journal} {\bibinfo
  {journal} {Nature Physics}\ }\textbf {\bibinfo {volume} {12}},\ \bibinfo
  {pages} {594} (\bibinfo {year} {2016})}\BibitemShut {NoStop}%
\bibitem [{\citenamefont {Cheal}\ and\ \citenamefont
  {Flanagan}(2010)}]{Cheal2010}%
  \BibitemOpen
  \bibfield  {author} {\bibinfo {author} {\bibfnamefont {B.}~\bibnamefont
  {Cheal}}\ and\ \bibinfo {author} {\bibfnamefont {K.~T.}\ \bibnamefont
  {Flanagan}},\ }\href {\doibase 10.1088/0954-3899/37/11/113101} {\bibfield
  {journal} {\bibinfo  {journal} {Journal of Physics G: Nuclear and Particle
  Physics}\ }\textbf {\bibinfo {volume} {37}},\ \bibinfo {pages} {113101}
  (\bibinfo {year} {2010})}\BibitemShut {NoStop}%
\bibitem [{\citenamefont {Blaum}\ \emph {et~al.}(2013)\citenamefont {Blaum},
  \citenamefont {Dilling},\ and\ \citenamefont
  {N\"{o}rtersh\"{a}user}}]{Blaum2013}%
  \BibitemOpen
  \bibfield  {author} {\bibinfo {author} {\bibfnamefont {K.}~\bibnamefont
  {Blaum}}, \bibinfo {author} {\bibfnamefont {J.}~\bibnamefont {Dilling}}, \
  and\ \bibinfo {author} {\bibfnamefont {W.}~\bibnamefont
  {N\"{o}rtersh\"{a}user}},\ }\href {\doibase
  10.1088/0031-8949/2013/T152/014017} {\bibfield  {journal} {\bibinfo
  {journal} {Physica Scripta}\ }\textbf {\bibinfo {volume} {T152}},\ \bibinfo
  {pages} {014017} (\bibinfo {year} {2013})}\BibitemShut {NoStop}%
\bibitem [{\citenamefont {Campbell}\ \emph {et~al.}(2016)\citenamefont
  {Campbell}, \citenamefont {Moore},\ and\ \citenamefont
  {Pearson}}]{Campbell2016}%
  \BibitemOpen
  \bibfield  {author} {\bibinfo {author} {\bibfnamefont {P.}~\bibnamefont
  {Campbell}}, \bibinfo {author} {\bibfnamefont {I.}~\bibnamefont {Moore}}, \
  and\ \bibinfo {author} {\bibfnamefont {M.}~\bibnamefont {Pearson}},\ }\href
  {\doibase http://dx.doi.org/10.1016/j.ppnp.2015.09.003} {\bibfield  {journal}
  {\bibinfo  {journal} {Progress in Particle and Nuclear Physics}\ }\textbf
  {\bibinfo {volume} {86}},\ \bibinfo {pages} {127 } (\bibinfo {year}
  {2016})}\BibitemShut {NoStop}%
\bibitem [{\citenamefont {Fedosseev}\ \emph {et~al.}(2012)\citenamefont
  {Fedosseev}, \citenamefont {Kudryavtsev},\ and\ \citenamefont
  {Mishin}}]{Fedosseev2012}%
  \BibitemOpen
  \bibfield  {author} {\bibinfo {author} {\bibfnamefont {V.~N.}\ \bibnamefont
  {Fedosseev}}, \bibinfo {author} {\bibfnamefont {Y.}~\bibnamefont
  {Kudryavtsev}}, \ and\ \bibinfo {author} {\bibfnamefont {V.~I.}\ \bibnamefont
  {Mishin}},\ }\href {\doibase 10.1088/0031-8949/85/05/058104} {\bibfield
  {journal} {\bibinfo  {journal} {Physica Scripta}\ }\textbf {\bibinfo {volume}
  {85}},\ \bibinfo {pages} {058104} (\bibinfo {year} {2012})}\BibitemShut
  {NoStop}%
\bibitem [{\citenamefont {De~Groote}\ \emph {et~al.}(2015)\citenamefont
  {De~Groote}, \citenamefont {Budin{\v{c}}evi{\'c}}, \citenamefont {Billowes},
  \citenamefont {Bissell}, \citenamefont {Cocolios}, \citenamefont
  {Farooq-Smith}, \citenamefont {Fedosseev}, \citenamefont {Flanagan},
  \citenamefont {Franchoo}, \citenamefont {Ruiz} \emph
  {et~al.}}]{de_Groote_2015}%
  \BibitemOpen
  \bibfield  {author} {\bibinfo {author} {\bibfnamefont {R.}~\bibnamefont
  {De~Groote}}, \bibinfo {author} {\bibfnamefont {I.}~\bibnamefont
  {Budin{\v{c}}evi{\'c}}}, \bibinfo {author} {\bibfnamefont {J.}~\bibnamefont
  {Billowes}}, \bibinfo {author} {\bibfnamefont {M.}~\bibnamefont {Bissell}},
  \bibinfo {author} {\bibfnamefont {T.}~\bibnamefont {Cocolios}}, \bibinfo
  {author} {\bibfnamefont {G.}~\bibnamefont {Farooq-Smith}}, \bibinfo {author}
  {\bibfnamefont {V.}~\bibnamefont {Fedosseev}}, \bibinfo {author}
  {\bibfnamefont {K.}~\bibnamefont {Flanagan}}, \bibinfo {author}
  {\bibfnamefont {S.}~\bibnamefont {Franchoo}}, \bibinfo {author}
  {\bibfnamefont {R.~G.}\ \bibnamefont {Ruiz}},  \emph {et~al.},\ }\href@noop
  {} {\bibfield  {journal} {\bibinfo  {journal} {Physical review letters}\
  }\textbf {\bibinfo {volume} {115}},\ \bibinfo {pages} {132501} (\bibinfo
  {year} {2015})}\BibitemShut {NoStop}%
\bibitem [{\citenamefont {Kudryavtsev}\ \emph {et~al.}(2013)\citenamefont
  {Kudryavtsev}, \citenamefont {Ferrer}, \citenamefont {Huyse}, \citenamefont
  {{Van den Bergh}},\ and\ \citenamefont {{Van Duppen}}}]{Kudryavtsev2013}%
  \BibitemOpen
  \bibfield  {author} {\bibinfo {author} {\bibfnamefont {Y.}~\bibnamefont
  {Kudryavtsev}}, \bibinfo {author} {\bibfnamefont {R.}~\bibnamefont {Ferrer}},
  \bibinfo {author} {\bibfnamefont {M.}~\bibnamefont {Huyse}}, \bibinfo
  {author} {\bibfnamefont {P.}~\bibnamefont {{Van den Bergh}}}, \ and\ \bibinfo
  {author} {\bibfnamefont {P.}~\bibnamefont {{Van Duppen}}},\ }\href {\doibase
  10.1016/j.nimb.2012.12.008} {\bibfield  {journal} {\bibinfo  {journal}
  {Nuclear Instruments and Methods in Physics Research Section B: Beam
  Interactions with Materials and Atoms}\ }\textbf {\bibinfo {volume} {297}},\
  \bibinfo {pages} {7} (\bibinfo {year} {2013})}\BibitemShut {NoStop}%
\bibitem [{\citenamefont {Raeder}\ \emph {et~al.}(2016)\citenamefont {Raeder},
  \citenamefont {Bastin}, \citenamefont {Block}, \citenamefont {Creemers},
  \citenamefont {Delahaye}, \citenamefont {Ferrer}, \citenamefont {Fléchard},
  \citenamefont {Franchoo}, \citenamefont {Ghys}, \citenamefont {Gaffney},
  \citenamefont {Granados}, \citenamefont {Heinke}, \citenamefont {Hijazi},
  \citenamefont {Huyse}, \citenamefont {Kron}, \citenamefont {Kudryavtsev},
  \citenamefont {Laatiaoui}, \citenamefont {Lecesne}, \citenamefont {Luton},
  \citenamefont {Moore}, \citenamefont {Martinez}, \citenamefont {Mogilevskiy},
  \citenamefont {Naubereit}, \citenamefont {Piot}, \citenamefont {Rothe},
  \citenamefont {Savajols}, \citenamefont {Sels}, \citenamefont {Sonnenschein},
  \citenamefont {Traykov}, \citenamefont {Beveren}, \citenamefont {den Bergh},
  \citenamefont {Duppen}, \citenamefont {Wendt},\ and\ \citenamefont
  {Zadvornaya}}]{Raeder2016}%
  \BibitemOpen
  \bibfield  {author} {\bibinfo {author} {\bibfnamefont {S.}~\bibnamefont
  {Raeder}}, \bibinfo {author} {\bibfnamefont {B.}~\bibnamefont {Bastin}},
  \bibinfo {author} {\bibfnamefont {M.}~\bibnamefont {Block}}, \bibinfo
  {author} {\bibfnamefont {P.}~\bibnamefont {Creemers}}, \bibinfo {author}
  {\bibfnamefont {P.}~\bibnamefont {Delahaye}}, \bibinfo {author}
  {\bibfnamefont {R.}~\bibnamefont {Ferrer}}, \bibinfo {author} {\bibfnamefont
  {X.}~\bibnamefont {Fléchard}}, \bibinfo {author} {\bibfnamefont
  {S.}~\bibnamefont {Franchoo}}, \bibinfo {author} {\bibfnamefont
  {L.}~\bibnamefont {Ghys}}, \bibinfo {author} {\bibfnamefont {L.}~\bibnamefont
  {Gaffney}}, \bibinfo {author} {\bibfnamefont {C.}~\bibnamefont {Granados}},
  \bibinfo {author} {\bibfnamefont {R.}~\bibnamefont {Heinke}}, \bibinfo
  {author} {\bibfnamefont {L.}~\bibnamefont {Hijazi}}, \bibinfo {author}
  {\bibfnamefont {M.}~\bibnamefont {Huyse}}, \bibinfo {author} {\bibfnamefont
  {T.}~\bibnamefont {Kron}}, \bibinfo {author} {\bibfnamefont {Y.}~\bibnamefont
  {Kudryavtsev}}, \bibinfo {author} {\bibfnamefont {M.}~\bibnamefont
  {Laatiaoui}}, \bibinfo {author} {\bibfnamefont {N.}~\bibnamefont {Lecesne}},
  \bibinfo {author} {\bibfnamefont {F.}~\bibnamefont {Luton}}, \bibinfo
  {author} {\bibfnamefont {I.}~\bibnamefont {Moore}}, \bibinfo {author}
  {\bibfnamefont {Y.}~\bibnamefont {Martinez}}, \bibinfo {author}
  {\bibfnamefont {E.}~\bibnamefont {Mogilevskiy}}, \bibinfo {author}
  {\bibfnamefont {P.}~\bibnamefont {Naubereit}}, \bibinfo {author}
  {\bibfnamefont {J.}~\bibnamefont {Piot}}, \bibinfo {author} {\bibfnamefont
  {S.}~\bibnamefont {Rothe}}, \bibinfo {author} {\bibfnamefont
  {H.}~\bibnamefont {Savajols}}, \bibinfo {author} {\bibfnamefont
  {S.}~\bibnamefont {Sels}}, \bibinfo {author} {\bibfnamefont {V.}~\bibnamefont
  {Sonnenschein}}, \bibinfo {author} {\bibfnamefont {E.}~\bibnamefont
  {Traykov}}, \bibinfo {author} {\bibfnamefont {C.~V.}\ \bibnamefont
  {Beveren}}, \bibinfo {author} {\bibfnamefont {P.~V.}\ \bibnamefont {den
  Bergh}}, \bibinfo {author} {\bibfnamefont {P.~V.}\ \bibnamefont {Duppen}},
  \bibinfo {author} {\bibfnamefont {K.}~\bibnamefont {Wendt}}, \ and\ \bibinfo
  {author} {\bibfnamefont {A.}~\bibnamefont {Zadvornaya}},\ }\href {\doibase
  http://dx.doi.org/10.1016/j.nimb.2015.12.014} {\bibfield  {journal} {\bibinfo
   {journal} {Nuclear Instruments and Methods in Physics Research Section B:
  Beam Interactions with Materials and Atoms}\ }\textbf {\bibinfo {volume}
  {376}},\ \bibinfo {pages} {382 } (\bibinfo {year} {2016})},\ \bibinfo {note}
  {proceedings of the \{XVIIth\} International Conference on Electromagnetic
  Isotope Separators and Related Topics (EMIS2015), Grand Rapids, MI, U.S.A.,
  11-15 May 2015}\BibitemShut {NoStop}%
\bibitem [{\citenamefont {Delone}\ and\ \citenamefont
  {Krainov}(1999)}]{Delone_1999}%
  \BibitemOpen
  \bibfield  {author} {\bibinfo {author} {\bibfnamefont {N.~B.}\ \bibnamefont
  {Delone}}\ and\ \bibinfo {author} {\bibfnamefont {V.~P.}\ \bibnamefont
  {Krainov}},\ }\href {\doibase 10.1070/pu1999v042n07abeh000557} {\bibfield
  {journal} {\bibinfo  {journal} {Physics-Uspekhi}\ }\textbf {\bibinfo {volume}
  {42}},\ \bibinfo {pages} {669} (\bibinfo {year} {1999})}\BibitemShut
  {NoStop}%
\bibitem [{\citenamefont {Kumekov}\ and\ \citenamefont
  {Perel}(1981)}]{kumekov1981dynamic}%
  \BibitemOpen
  \bibfield  {author} {\bibinfo {author} {\bibfnamefont {S.}~\bibnamefont
  {Kumekov}}\ and\ \bibinfo {author} {\bibfnamefont {V.}~\bibnamefont
  {Perel}},\ }\href@noop {} {\bibfield  {journal} {\bibinfo  {journal} {Zh.
  Eksp. Teor. Fiz}\ }\textbf {\bibinfo {volume} {81}},\ \bibinfo {pages} {1693}
  (\bibinfo {year} {1981})}\BibitemShut {NoStop}%
\bibitem [{\citenamefont {Knight}\ and\ \citenamefont
  {Lauder}(1990)}]{Knight1990}%
  \BibitemOpen
  \bibfield  {author} {\bibinfo {author} {\bibfnamefont {P.~L.}\ \bibnamefont
  {Knight}}\ and\ \bibinfo {author} {\bibfnamefont {M.~A.}\ \bibnamefont
  {Lauder}},\ }\href@noop {} {\bibfield  {journal} {\bibinfo  {journal}
  {Physics Reports}\ }\textbf {\bibinfo {volume} {190}},\ \bibinfo {pages} {1}
  (\bibinfo {year} {1990})}\BibitemShut {NoStop}%
\bibitem [{\citenamefont {Dai}\ and\ \citenamefont
  {Lambropoulos}(1987)}]{Dai1987}%
  \BibitemOpen
  \bibfield  {author} {\bibinfo {author} {\bibfnamefont {B.~N.}\ \bibnamefont
  {Dai}}\ and\ \bibinfo {author} {\bibfnamefont {P.}~\bibnamefont
  {Lambropoulos}},\ }\href@noop {} {\bibfield  {journal} {\bibinfo  {journal}
  {Physical Review A}\ }\textbf {\bibinfo {volume} {36}},\ \bibinfo {pages}
  {5205} (\bibinfo {year} {1987})}\BibitemShut {NoStop}%
\bibitem [{\citenamefont {Nakajima}\ \emph {et~al.}(1994)\citenamefont
  {Nakajima}, \citenamefont {Elk}, \citenamefont {Zhang},\ and\ \citenamefont
  {Lambropoulos}}]{Nakajima1994}%
  \BibitemOpen
  \bibfield  {author} {\bibinfo {author} {\bibfnamefont {T.}~\bibnamefont
  {Nakajima}}, \bibinfo {author} {\bibfnamefont {M.}~\bibnamefont {Elk}},
  \bibinfo {author} {\bibfnamefont {J.}~\bibnamefont {Zhang}}, \ and\ \bibinfo
  {author} {\bibfnamefont {P.}~\bibnamefont {Lambropoulos}},\ }\href@noop {}
  {\bibfield  {journal} {\bibinfo  {journal} {Physical Review A}\ }\textbf
  {\bibinfo {volume} {50}},\ \bibinfo {pages} {915} (\bibinfo {year}
  {1994})}\BibitemShut {NoStop}%
\bibitem [{\citenamefont {Yatsenko}\ \emph {et~al.}(1997)\citenamefont
  {Yatsenko}, \citenamefont {Unanyan}, \citenamefont {Bergmann}, \citenamefont
  {Halfmann},\ and\ \citenamefont {Shore}}]{Yatsenko1997}%
  \BibitemOpen
  \bibfield  {author} {\bibinfo {author} {\bibfnamefont {L.~P.}\ \bibnamefont
  {Yatsenko}}, \bibinfo {author} {\bibfnamefont {R.~G.}\ \bibnamefont
  {Unanyan}}, \bibinfo {author} {\bibfnamefont {K.}~\bibnamefont {Bergmann}},
  \bibinfo {author} {\bibfnamefont {T.}~\bibnamefont {Halfmann}}, \ and\
  \bibinfo {author} {\bibfnamefont {B.~W.}\ \bibnamefont {Shore}},\ }\href@noop
  {} {\bibfield  {journal} {\bibinfo  {journal} {Optics Communications}\
  }\textbf {\bibinfo {volume} {135}},\ \bibinfo {pages} {406} (\bibinfo {year}
  {1997})}\BibitemShut {NoStop}%
\bibitem [{\citenamefont {Yatsenko}\ \emph {et~al.}(1999)\citenamefont
  {Yatsenko}, \citenamefont {Halfmann}, \citenamefont {Shore},\ and\
  \citenamefont {Bergmann}}]{Yatsenko1999}%
  \BibitemOpen
  \bibfield  {author} {\bibinfo {author} {\bibfnamefont {L.}~\bibnamefont
  {Yatsenko}}, \bibinfo {author} {\bibfnamefont {T.}~\bibnamefont {Halfmann}},
  \bibinfo {author} {\bibfnamefont {B.}~\bibnamefont {Shore}}, \ and\ \bibinfo
  {author} {\bibfnamefont {K.}~\bibnamefont {Bergmann}},\ }\href {\doibase
  10.1103/PhysRevA.59.2926} {\bibfield  {journal} {\bibinfo  {journal}
  {Physical Review A}\ }\textbf {\bibinfo {volume} {59}},\ \bibinfo {pages}
  {2926} (\bibinfo {year} {1999})}\BibitemShut {NoStop}%
\bibitem [{\citenamefont {Citron}\ \emph {et~al.}(1977)\citenamefont {Citron},
  \citenamefont {Gray}, \citenamefont {Gabel},\ and\ \citenamefont
  {Stroud}}]{Citron_1977}%
  \BibitemOpen
  \bibfield  {author} {\bibinfo {author} {\bibfnamefont {M.~L.}\ \bibnamefont
  {Citron}}, \bibinfo {author} {\bibfnamefont {H.~R.}\ \bibnamefont {Gray}},
  \bibinfo {author} {\bibfnamefont {C.~W.}\ \bibnamefont {Gabel}}, \ and\
  \bibinfo {author} {\bibfnamefont {C.~R.}\ \bibnamefont {Stroud}},\ }\href
  {\doibase 10.1103/physreva.16.1507} {\bibfield  {journal} {\bibinfo
  {journal} {Phys. Rev. A}\ }\textbf {\bibinfo {volume} {16}},\ \bibinfo
  {pages} {1507} (\bibinfo {year} {1977})}\BibitemShut {NoStop}%
\bibitem [{\citenamefont {Vitanov}(2001)}]{Vitanov2001}%
  \BibitemOpen
  \bibfield  {author} {\bibinfo {author} {\bibfnamefont {N.}~\bibnamefont
  {Vitanov}},\ }\href
  {http://linkinghub.elsevier.com/retrieve/pii/S003040180101495X} {\bibfield
  {journal} {\bibinfo  {journal} {Optics Communications}\ }\textbf {\bibinfo
  {volume} {199}},\ \bibinfo {pages} {117} (\bibinfo {year}
  {2001})}\BibitemShut {NoStop}%
\bibitem [{\citenamefont {Boradjiev}\ and\ \citenamefont
  {Vitanov}(2013)}]{Boradjiev2013}%
  \BibitemOpen
  \bibfield  {author} {\bibinfo {author} {\bibfnamefont {I.~I.}\ \bibnamefont
  {Boradjiev}}\ and\ \bibinfo {author} {\bibfnamefont {N.~V.}\ \bibnamefont
  {Vitanov}},\ }\href {\doibase 10.1016/j.optcom.2012.09.040} {\bibfield
  {journal} {\bibinfo  {journal} {Optics Communications}\ }\textbf {\bibinfo
  {volume} {288}},\ \bibinfo {pages} {91} (\bibinfo {year} {2013})}\BibitemShut
  {NoStop}%
\bibitem [{\citenamefont {Kono}\ and\ \citenamefont
  {Hattori}(1982)}]{Kono1982}%
  \BibitemOpen
  \bibfield  {author} {\bibinfo {author} {\bibfnamefont {A.}~\bibnamefont
  {Kono}}\ and\ \bibinfo {author} {\bibfnamefont {S.}~\bibnamefont {Hattori}},\
  }\href {\doibase http://dx.doi.org/10.1016/0022-4073(82)90003-6} {\bibfield
  {journal} {\bibinfo  {journal} {Journal of Quantitative Spectroscopy and
  Radiative Transfer}\ }\textbf {\bibinfo {volume} {28}},\ \bibinfo {pages}
  {383 } (\bibinfo {year} {1982})}\BibitemShut {NoStop}%
\bibitem [{\citenamefont {Kessler}\ \emph {et~al.}(2008)\citenamefont
  {Kessler}, \citenamefont {Moore}, \citenamefont {Kudryavtsev}, \citenamefont
  {Peräjärvi}, \citenamefont {Popov}, \citenamefont {Ronkanen}, \citenamefont
  {Sonoda}, \citenamefont {Tordoff}, \citenamefont {Wendt},\ and\ \citenamefont
  {Äystö}}]{Kessler2008}%
  \BibitemOpen
  \bibfield  {author} {\bibinfo {author} {\bibfnamefont {T.}~\bibnamefont
  {Kessler}}, \bibinfo {author} {\bibfnamefont {I.}~\bibnamefont {Moore}},
  \bibinfo {author} {\bibfnamefont {Y.}~\bibnamefont {Kudryavtsev}}, \bibinfo
  {author} {\bibfnamefont {K.}~\bibnamefont {Peräjärvi}}, \bibinfo {author}
  {\bibfnamefont {A.}~\bibnamefont {Popov}}, \bibinfo {author} {\bibfnamefont
  {P.}~\bibnamefont {Ronkanen}}, \bibinfo {author} {\bibfnamefont
  {T.}~\bibnamefont {Sonoda}}, \bibinfo {author} {\bibfnamefont
  {B.}~\bibnamefont {Tordoff}}, \bibinfo {author} {\bibfnamefont
  {K.}~\bibnamefont {Wendt}}, \ and\ \bibinfo {author} {\bibfnamefont
  {J.}~\bibnamefont {Äystö}},\ }\href {\doibase
  http://dx.doi.org/10.1016/j.nimb.2007.11.076} {\bibfield  {journal} {\bibinfo
   {journal} {Nuclear Instruments and Methods in Physics Research Section B:
  Beam Interactions with Materials and Atoms}\ }\textbf {\bibinfo {volume}
  {266}},\ \bibinfo {pages} {681 } (\bibinfo {year} {2008})}\BibitemShut
  {NoStop}%
\bibitem [{\citenamefont {{Sonnenschein}}(2015)}]{Sonnenschein2015}%
  \BibitemOpen
  \bibfield  {author} {\bibinfo {author} {\bibfnamefont {V.}~\bibnamefont
  {{Sonnenschein}}},\ }\emph {\bibinfo {title} {{Laser Developments and High
  Resolution Resonance Ionization Spectroscopy of Actinide Elements}}},\
  \href@noop {} {Ph.D. thesis} (\bibinfo {year} {2015})\BibitemShut {NoStop}%
\bibitem [{\citenamefont {Flanagan}\ \emph {et~al.}(2013)\citenamefont
  {Flanagan}, \citenamefont {Lynch}, \citenamefont {Billowes}, \citenamefont
  {Bissell}, \citenamefont {Budin\v{c}evi\'{c}}, \citenamefont {Cocolios},
  \citenamefont {de~Groote}, \citenamefont {{De Schepper}}, \citenamefont
  {Fedosseev}, \citenamefont {Franchoo}, \citenamefont {{Garcia Ruiz}},
  \citenamefont {Heylen}, \citenamefont {Marsh}, \citenamefont {Neyens},
  \citenamefont {Procter}, \citenamefont {Rossel}, \citenamefont {Rothe},
  \citenamefont {Strashnov}, \citenamefont {Stroke},\ and\ \citenamefont
  {Wendt}}]{Flanagan2013}%
  \BibitemOpen
  \bibfield  {author} {\bibinfo {author} {\bibfnamefont {K.~T.}\ \bibnamefont
  {Flanagan}}, \bibinfo {author} {\bibfnamefont {K.~M.}\ \bibnamefont {Lynch}},
  \bibinfo {author} {\bibfnamefont {J.}~\bibnamefont {Billowes}}, \bibinfo
  {author} {\bibfnamefont {M.~L.}\ \bibnamefont {Bissell}}, \bibinfo {author}
  {\bibfnamefont {I.}~\bibnamefont {Budin\v{c}evi\'{c}}}, \bibinfo {author}
  {\bibfnamefont {T.~E.}\ \bibnamefont {Cocolios}}, \bibinfo {author}
  {\bibfnamefont {R.~P.}\ \bibnamefont {de~Groote}}, \bibinfo {author}
  {\bibfnamefont {S.}~\bibnamefont {{De Schepper}}}, \bibinfo {author}
  {\bibfnamefont {V.~N.}\ \bibnamefont {Fedosseev}}, \bibinfo {author}
  {\bibfnamefont {S.}~\bibnamefont {Franchoo}}, \bibinfo {author}
  {\bibfnamefont {R.~F.}\ \bibnamefont {{Garcia Ruiz}}}, \bibinfo {author}
  {\bibfnamefont {H.}~\bibnamefont {Heylen}}, \bibinfo {author} {\bibfnamefont
  {B.~a.}\ \bibnamefont {Marsh}}, \bibinfo {author} {\bibfnamefont
  {G.}~\bibnamefont {Neyens}}, \bibinfo {author} {\bibfnamefont {T.~J.}\
  \bibnamefont {Procter}}, \bibinfo {author} {\bibfnamefont {R.~E.}\
  \bibnamefont {Rossel}}, \bibinfo {author} {\bibfnamefont {S.}~\bibnamefont
  {Rothe}}, \bibinfo {author} {\bibfnamefont {I.}~\bibnamefont {Strashnov}},
  \bibinfo {author} {\bibfnamefont {H.~H.}\ \bibnamefont {Stroke}}, \ and\
  \bibinfo {author} {\bibfnamefont {K.~D.~A.}\ \bibnamefont {Wendt}},\ }\href
  {\doibase 10.1103/PhysRevLett.111.212501} {\bibfield  {journal} {\bibinfo
  {journal} {Physical Review Letters}\ }\textbf {\bibinfo {volume} {111}},\
  \bibinfo {pages} {212501} (\bibinfo {year} {2013})}\BibitemShut {NoStop}%
\bibitem [{\citenamefont {Duong}\ \emph {et~al.}(1987)\citenamefont {Duong},
  \citenamefont {Juncar}, \citenamefont {Liberman}, \citenamefont {Mueller},
  \citenamefont {Neugart}, \citenamefont {Otten}, \citenamefont {Peuse},
  \citenamefont {Pinard}, \citenamefont {Stroke}, \citenamefont {Thibault},
  \citenamefont {Touchard}, \citenamefont {Vialle}, \citenamefont {Wendt},\
  and\ \citenamefont {Collaboration}}]{Duong1987}%
  \BibitemOpen
  \bibfield  {author} {\bibinfo {author} {\bibfnamefont {H.~T.}\ \bibnamefont
  {Duong}}, \bibinfo {author} {\bibfnamefont {P.}~\bibnamefont {Juncar}},
  \bibinfo {author} {\bibfnamefont {S.}~\bibnamefont {Liberman}}, \bibinfo
  {author} {\bibfnamefont {A.~C.}\ \bibnamefont {Mueller}}, \bibinfo {author}
  {\bibfnamefont {R.}~\bibnamefont {Neugart}}, \bibinfo {author} {\bibfnamefont
  {E.~W.}\ \bibnamefont {Otten}}, \bibinfo {author} {\bibfnamefont
  {B.}~\bibnamefont {Peuse}}, \bibinfo {author} {\bibfnamefont
  {J.}~\bibnamefont {Pinard}}, \bibinfo {author} {\bibfnamefont {H.~H.}\
  \bibnamefont {Stroke}}, \bibinfo {author} {\bibfnamefont {C.}~\bibnamefont
  {Thibault}}, \bibinfo {author} {\bibfnamefont {F.}~\bibnamefont {Touchard}},
  \bibinfo {author} {\bibfnamefont {J.~L.}\ \bibnamefont {Vialle}}, \bibinfo
  {author} {\bibfnamefont {K.}~\bibnamefont {Wendt}}, \ and\ \bibinfo {author}
  {\bibfnamefont {I.}~\bibnamefont {Collaboration}},\ }\href
  {http://stacks.iop.org/0295-5075/3/i=2/a=008} {\bibfield  {journal} {\bibinfo
   {journal} {EPL (Europhysics Letters)}\ }\textbf {\bibinfo {volume} {3}},\
  \bibinfo {pages} {175} (\bibinfo {year} {1987})}\BibitemShut {NoStop}%
\bibitem [{\citenamefont {Procter}\ \emph {et~al.}(2012)\citenamefont
  {Procter}, \citenamefont {Aghaei-Khozani}, \citenamefont {Billowes},
  \citenamefont {Bissell}, \citenamefont {Blanc}, \citenamefont {Cheal},
  \citenamefont {Cocolios}, \citenamefont {Flanagan}, \citenamefont {Hori},
  \citenamefont {Kobayashi}, \citenamefont {Lunney}, \citenamefont {Lynch},
  \citenamefont {Marsh}, \citenamefont {Neyens}, \citenamefont {Papuga},
  \citenamefont {Rajabali}, \citenamefont {Rothe}, \citenamefont {Simpson},
  \citenamefont {Smith}, \citenamefont {Stroke}, \citenamefont
  {Vanderheijden},\ and\ \citenamefont {Wendt}}]{Procter2012}%
  \BibitemOpen
  \bibfield  {author} {\bibinfo {author} {\bibfnamefont {T.~J.}\ \bibnamefont
  {Procter}}, \bibinfo {author} {\bibfnamefont {H.}~\bibnamefont
  {Aghaei-Khozani}}, \bibinfo {author} {\bibfnamefont {J.}~\bibnamefont
  {Billowes}}, \bibinfo {author} {\bibfnamefont {M.~L.}\ \bibnamefont
  {Bissell}}, \bibinfo {author} {\bibfnamefont {F.~L.}\ \bibnamefont {Blanc}},
  \bibinfo {author} {\bibfnamefont {B.}~\bibnamefont {Cheal}}, \bibinfo
  {author} {\bibfnamefont {T.~E.}\ \bibnamefont {Cocolios}}, \bibinfo {author}
  {\bibfnamefont {K.~T.}\ \bibnamefont {Flanagan}}, \bibinfo {author}
  {\bibfnamefont {H.}~\bibnamefont {Hori}}, \bibinfo {author} {\bibfnamefont
  {T.}~\bibnamefont {Kobayashi}}, \bibinfo {author} {\bibfnamefont
  {D.}~\bibnamefont {Lunney}}, \bibinfo {author} {\bibfnamefont {K.~M.}\
  \bibnamefont {Lynch}}, \bibinfo {author} {\bibfnamefont {B.~a.}\ \bibnamefont
  {Marsh}}, \bibinfo {author} {\bibfnamefont {G.}~\bibnamefont {Neyens}},
  \bibinfo {author} {\bibfnamefont {J.}~\bibnamefont {Papuga}}, \bibinfo
  {author} {\bibfnamefont {M.~M.}\ \bibnamefont {Rajabali}}, \bibinfo {author}
  {\bibfnamefont {S.}~\bibnamefont {Rothe}}, \bibinfo {author} {\bibfnamefont
  {G.}~\bibnamefont {Simpson}}, \bibinfo {author} {\bibfnamefont {A.~J.}\
  \bibnamefont {Smith}}, \bibinfo {author} {\bibfnamefont {H.~H.}\ \bibnamefont
  {Stroke}}, \bibinfo {author} {\bibfnamefont {W.}~\bibnamefont
  {Vanderheijden}}, \ and\ \bibinfo {author} {\bibfnamefont {K.}~\bibnamefont
  {Wendt}},\ }\href {\doibase 10.1088/1742-6596/381/1/012070} {\bibfield
  {journal} {\bibinfo  {journal} {Journal of Physics: Conference Series}\
  }\textbf {\bibinfo {volume} {381}},\ \bibinfo {pages} {012070} (\bibinfo
  {year} {2012})}\BibitemShut {NoStop}%
\bibitem [{\citenamefont {Ambartzumian}\ \emph {et~al.}(1976)\citenamefont
  {Ambartzumian}, \citenamefont {Furzikov}, \citenamefont {Letokhov},\ and\
  \citenamefont {Puretsky}}]{Ambartzumian1976}%
  \BibitemOpen
  \bibfield  {author} {\bibinfo {author} {\bibfnamefont {R.~V.}\ \bibnamefont
  {Ambartzumian}}, \bibinfo {author} {\bibfnamefont {N.~P.}\ \bibnamefont
  {Furzikov}}, \bibinfo {author} {\bibfnamefont {V.~S.}\ \bibnamefont
  {Letokhov}}, \ and\ \bibinfo {author} {\bibfnamefont {A.~A.}\ \bibnamefont
  {Puretsky}},\ }\href {\doibase 10.1007/BF00900460} {\bibfield  {journal}
  {\bibinfo  {journal} {Applied physics}\ }\textbf {\bibinfo {volume} {9}},\
  \bibinfo {pages} {335} (\bibinfo {year} {1976})}\BibitemShut {NoStop}%
\bibitem [{\citenamefont {Gilbert}\ \emph {et~al.}(1984)\citenamefont
  {Gilbert}, \citenamefont {Noecker},\ and\ \citenamefont
  {Wieman}}]{Gilbert1984}%
  \BibitemOpen
  \bibfield  {author} {\bibinfo {author} {\bibfnamefont {S.~L.}\ \bibnamefont
  {Gilbert}}, \bibinfo {author} {\bibfnamefont {M.~C.}\ \bibnamefont
  {Noecker}}, \ and\ \bibinfo {author} {\bibfnamefont {C.~E.}\ \bibnamefont
  {Wieman}},\ }\href {\doibase 10.1103/PhysRevA.29.3150} {\bibfield  {journal}
  {\bibinfo  {journal} {Phys. Rev. A}\ }\textbf {\bibinfo {volume} {29}},\
  \bibinfo {pages} {3150} (\bibinfo {year} {1984})}\BibitemShut {NoStop}%
\end{thebibliography}%

\end{document}